\documentclass[11pt,preprint2]{aastex}

\newcommand{\Ha}{H$\alpha$}

\newcommand{\Hb}{H$\beta$}
\newcommand{\htwo}{H$_{2}$}

\newcommand{\etal}{et~al.}
\newcommand{\kms}{km~s{$^{-1}$}}

\newcommand{\cmq}{cm$^{-3}$}

\slugcomment{Submitted to the Astronomical Journal}

\shorttitle{Unraveling the Helix Nebula}
\shortauthors{O'Dell, McCullough, and Meixner}

\begin{document}


\title{Unraveling the Helix Nebula: Its Structure and Knots
\footnote{
Based in part on observations with the NASA/ESA Hubble Space Telescope,
obtained at the Space Telescope Science Institute, which is operated by
the Association of Universities for Research in Astronomy, Inc., under
NASA Contract No. NAS 5-26555.}
\footnote{
Based in part on observations obtained at the Cerro Tololo Interamerican Observatory,
which is operated by
the Association of Universities for Research in Astronomy, Inc., under
a Cooperative Agreement with the National Science Foundation.}}


\author{C. R. O'Dell}
\affil{Department of Physics and Astronomy, Vanderbilt University,
Box 1807-B, Nashville, TN 37204}
\email{cr.odell@vanderbilt.edu}
\author{Peter R. McCullough}
\affil{Space Telescope Science Institute, 3700 San Martin Drive, Baltimore,
MD 21218}
\and
\author{Margaret Meixner}
\affil{Space Telescope Science Institute, 3700 San Martin Drive, Baltimore,
MD 21218}

\begin{abstract}
Through HST imaging of the inner part of the main-ring of the Helix Nebula together with
CTIO 4-m images of the fainter outer parts, we have an unprecedented-quality view of 
the nearest bright planetary nebula. These images have allowed
determination that the main-ring of the nebula is composed of an inner-disk of about 499\arcsec\
diameter (0.52 pc) surrounded by an outer-ring (in reality a torus) of 742\arcsec\ diameter (0.77 pc)
whose plane is highly inclined to the plane of the disk. This outer-ring is surrounded by
an outermost-ring of 1500\arcsec\ (1.76 pc) diameter which is flattened on the side colliding with 
the ambient interstellar medium.  The inner-disk has an extended distribution of low density
gas along its rotational axis of symmetry and the disk is optically thick to ionizing radiation,
as is the outer-ring.

Published radial velocities of the knots provides support for the two-component structure of
the main-ring of the nebula and to the idea that the knots found there are expanding along with the
nebular material from which it recently originated.  These velocities indicate a spatial expansion 
velocity of the inner-disk of 40 \kms\ and 32 \kms\ for the outer-ring, which yields expansion ages 
of 6,560 yrs and 12,100 yrs respectively.

The outermost-ring may be partially ionized through scattered recombination
continuum from the inner parts of the nebula, but shocks certainly are occurring in it. 
This outermost-ring probably represents a third period of mass-loss by the central star.
There is one compact, outer object which is unexplained, showing shock structures indicating a different
orientation of the gas flow from that of the nebula.  

There is a change in the morphology of the knots as a function of the distance from the local ionization
front. This supports a scenario in which the knots are formed in or near the ionization front and are 
then sculpted by the stellar radiation from the central star as the ionization front advances beyond them.

\end{abstract}


\section{INTRODUCTION}

\begin{figure}
\plotone{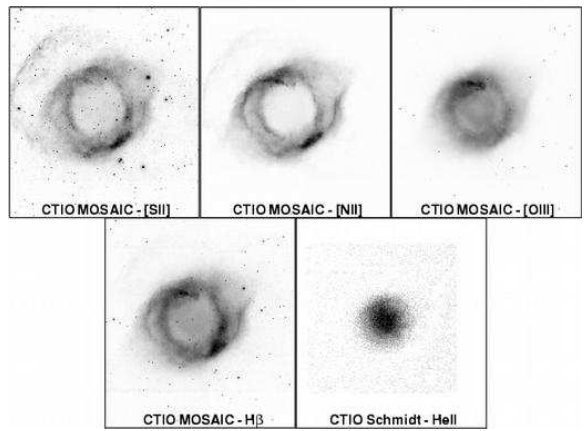}
\caption{
{These negative depictions of CTIO images covers a field of 1500\arcsec x 1653\arcsec with the vertical axes pointed
towards PA=0.9\arcdeg. The grey-level was adjusted so that saturation was reached only in a single brightest region.
The 2.64\arcsec\ resolution of these depictions is set by binning the MOSAIC data into ten pixels and scaling the Schmidt
data to the same value. The considerable difference of appearance of the main-ring in different ions is discussed in
\S\ 3 and \S\ 4.}
} \label{fig-1}
\end{figure}

\begin{figure}
\plotone{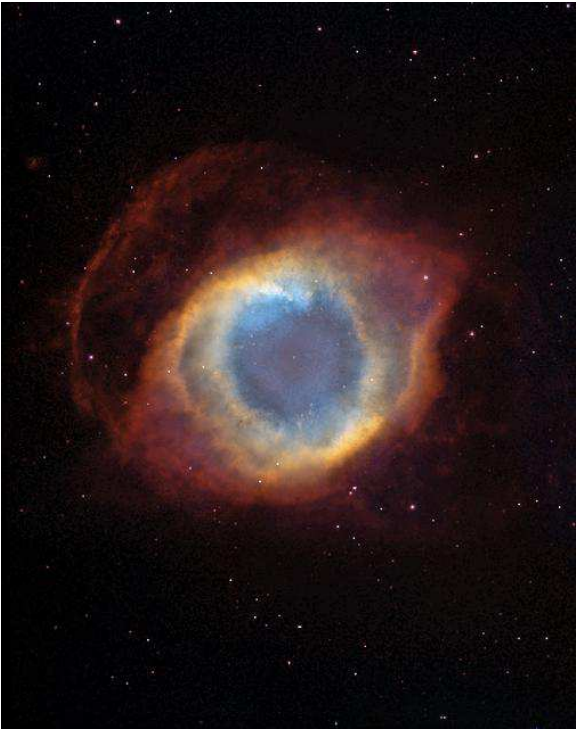}
\caption{
{This composite image of the CTIO and HST observations was made with \Ha +[N~II] as red, an average of \Ha +[N~II] and [O~III] as green,  and [O~III] as blue. In each
color the image was made by combining the ACS observations of the inner nebula with the MOSAIC images of the outer nebula
as described in \S\ 2.3.
The scale of the original composite image is 0.264\arcsec /pixel which was degraded to 0.533\arcsec /pixel and printed at 300 pixels per inch in order to fit
the printing restrictions of the journal, the full resolution version is available as a 165 MB file, helixhocaclean.tif, at ftp://cauldron.stsci.edu/world/outbox/levay/helix/.}
} \label{fig-2}
\end{figure}

The Helix Nebula (NGC 7293) occupies a special place in the study of the planetary nebulae (PN)
because it is the closest bright member of this important class of objects. It is difficult
to observe, when compared with many other bright PN, because its surface brightness is relatively low due to the nebula's large
angular size. Studies of the appearance in optical emission lines with ground-based telescopes
and the Hubble Space Telescope (HST) has produced a series of spectacular new images 
Meaburn \etal\ 1992 (henceforth M1992),
O'Dell \&\ Handron 1996 (henceforth OH1996), and O'Dell \& Burkert 1997, (henceforth OB1997)
revealing the ubiquity and detail within the knots that populate the object (Zanstra 1955, Vorontzov-Velyaminov 1968).
A recent study of several nearby PN (O'Dell \etal\ 2002, henceforth O2002) indicate that these knots are
common features in PN and may therefore, play an important role in the evolution of these 
ejected stellar envelopes.

Even though the HST images (including the new ones reported upon in this study) provide
un-excelled resolution, these images do not cover the entire nebula, which means that there
is the need to continue to image the object from the ground with an increasingly improved
resolution and photometric accuracy. The most useful photometric ground observations are
those of O'Dell (1998, henceforth O1998) in the \Hb\, [O~III], and HeII lines, although
his resolution was determined by the 2.3\arcsec\ pixels employed. Images in other bands
have significantly added to our knowledge of the object. Speck \etal\ (2002, henceforth S2002)
have imaged the entire main-ring of the nebula in the \htwo\ 2.12 \micron\ line at a resolution
of 2\arcsec, while Meixner \etal\ (2004) have imaged a fraction of the southeast main-ring
region in the same line with a resolution of 0.2\arcsec. 
Although the \htwo\ study of Cox \etal\ (1998) has a resolution of only 6\arcsec\, it includes
multiple lines and was able to establish that the \htwo\ has a temperature of about 900 K.
The entire main-ring has been mapped in
CO emission with resolutions as good as 31\arcsec\ (Young \etal\ 1999 , henceforth Y1999). 
Huggins \etal\ (2002, henceforth H2002) has imaged in CO the region of the bright knot 378-801
with a beam of 7.9\arcsec x 3.8\arcsec.
Rodr\'{\i}quez, Goss, \&\ Williams (2002) 
has mapped the main-ring in the 21-cm line of HI with a beam of 
54.3\arcsec x 39.3\arcsec.
Although the radio observations fall far below the spatial resolution of the optical observations,
the multiplicity of sources within their beams that are revealed by the superb velocity
resolution gives indications of the finer scale structure that must exist.

The Helix Nebula has usually been assigned as a member of the bipolar class of PN and the commonly accepted model is a thick
disk of ionized gas that is optically thick along its equator and may be optically
thin to ionizing flux along the symmetry axis (Meaburn \etal\ 1998, henceforth M1998; O1998,
Henry,  Kwitter, \&\ Dufour 1999, henceforth H1999).  In this simple disk model the 
symmetry axis is pointed about 30\arcdeg\ with respect
to a line of sight to the observer. In this nearly face-on view of the disk, we 
see that its inner region is occupied by a core of He$^{++}$ emission, whose lack of
ions with high emissivity give the nebula a superficial appearance of having a 
central cavity (O1998). This single disk model must now be superseded by a more complex model,
as described in \S\ 3.1, \S\ 4.1 and \S\ 4.2.
The existence of plumes extending from the main-ring of emission to
the northwest and southeast indicate that the structure is not a simple disk, rather, there
must be extended material above and below the disk.
The presence of [S~II] emission even
across the region of the central star indicates that the nebula is likely to be optically
thick in most directions, since this emission usually arises only from an ionization front (although
shock excitation is also possible in some regions, \S\ 4.3). The knots begin to be observed
at about the same distance from the central star that the HeII core gives way to the lower
ionization H$^{+}$+He$^{+}$ zone that produces the dominant [O~III] emission.

\begin{figure}
\plotone{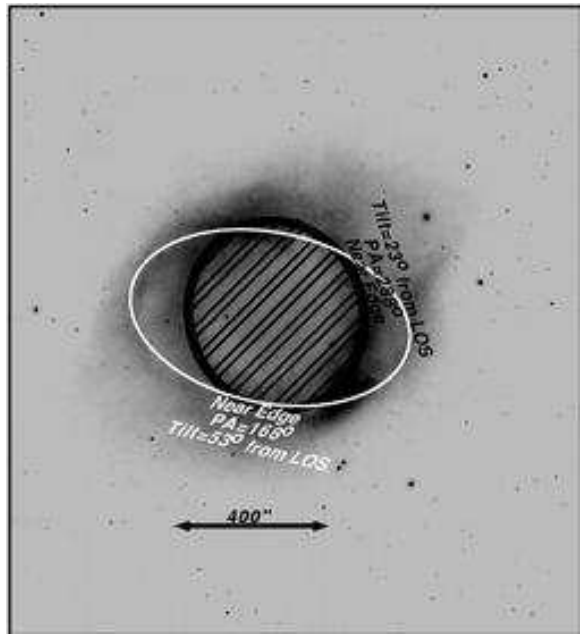}
\caption{
{This negative depiction image of the Helix Nebula in the CTIO \Hb\ image has drawn upon it two ellipses which
define the inner-ring and the outer-rings within the main-ring of the nebula. The three dimensional properties of 
these objects-when interpreted as circles seen at oblique angles are-labeled. Dark lines indicate the inner-ring
and white lines the outer-ring.}
} \label{fig-3}
\end{figure}

Throughout this paper we use a distance of 213 parsecs, which is based on
the trigonometric parallax of the central star (Harris \etal\ 1997). This yields
a scale of 1\arcsec\ = 3.19 x 10$^{15}$ cm and the 5.45\arcmin\ semimajor axis of the nebula (O2002)
is then 0.33 pc.

It is necessary to comment on the nomenclature used in this paper since different
and sometimes conflicting designations are used for the same features in different papers.
By the main-ring of the nebula we mean the ring-shaped structure seen on images of the nebula.
Within this structure there are loop structures. As we'll see in \S\ 3, some of these
loops are related and form continuous structures called the inner-disk and outer-ring. The 
fainter extensions to the northwest and southeast will be called the plumes. The outermost
structures to the east and west  of the nebula will be called the outer-nebula.
The compact condensations will be called knots, with the projected regions immediately
beyond them (away from the central star) being called their tails. The fine-scale
radial structures seen on the outside of the main body will be called radial-rays. Whenever a
small feature needs specific designation, we'll use the coordinate based system
introduced in OB1997, thus avoiding the confusion of a discovery-time based system.
We give figures labeling features at decreasing surface brightness and increasing size in
Figure 3, Figure 6, and Figure 17.

We describe the new high resolution HST and broader field of view (FOV) ground-based observations in \S\ 2. 
In \S\ 3 we describe the analysis of these new observations and discuss these
results in \S\ 4. What we find is that the Helix Nebula is composed of an inner-disk of material,
surrounded by a  lower ionization outer-ring that is almost perpendicular to the inner-disk. This interpretation
is strengthened by its explanation of the complex radial velocity pattern described by radio and
optical observations. There is an outer structure called the outermost-ring whose form suggests that
it has the same orientation as the outer-ring. The northeast portion of this outer-ring 
is compressed by collision with the interstellar medium (ISM). There is evidence for
continued outflow along an axis perpendicular to the inner-disk and there is an east-outer feature that
appears to be the result of a collimated outflow shocking against the ISM. Several other shock-form
objects are also seen, although their origin and nature are unexplained. We trace the change of 
morphological appearance of the well-known knots and determine that these objects are first seen
soon after the passage of the nebula's ionization front as broad structures with significant amounts
of material in their outer regions. Those objects lying farther behind the ionization front have
evolved in form to be well-defined structures with mini-ionization fronts giving them the bright
cusps facing the central star and less material in their outward tail regions.

\begin{figure}
\plotone{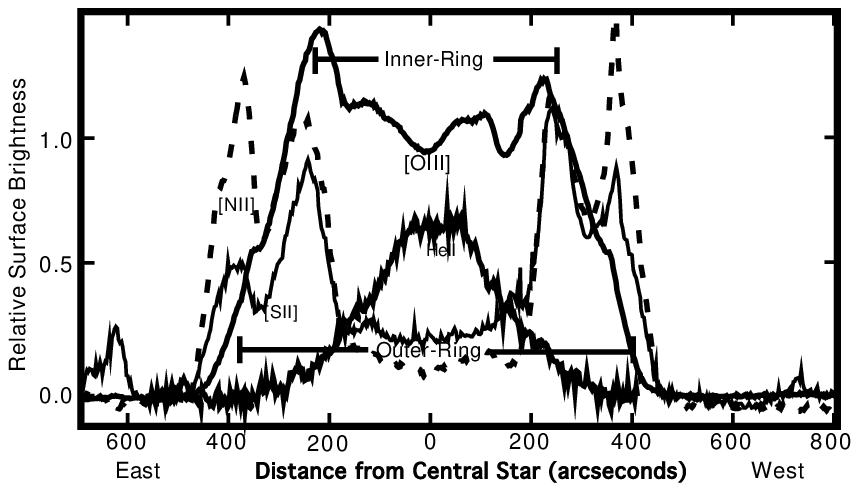}
\caption{
{The brightness profile of an east-west sample 57.5\arcsec\ wide across the middle of the CTIO images
is shown, with signal from different ions scaled into arbitrary intensity units for clarity. Features associated with the inner-disk and the
outer-ring are labeled. Features beyond 500\arcsec\ are associated with the low ionization outer-nebula.}
} \label{fig-4}
\end{figure}

\section{OBSERVATIONS}

We draw on several related data sets for this study, the first being a mosaic of images (McCullough et~al. 2002) made with the
ACS WFC instrument (Pavlovsky 2003) of the HST, the second a set of Cerro Tololo Interamerican
Observatory (CTIO) 4-m ground-based observations, and the third is a limited set of previously published observations with
with the Curtis-Schmidt telescope at CTIO. In this section we describe the first two sets and the third has already
been described in a publication (O1998).

\subsection{A Mosaic of HST ACS Images in Two Filters}

The passage of the Earth through the debris stream that produces the Leonids meteor shower was thought to 
present a risk to the HST for the several years when the Earth passed close to the center of the stream.
During these passages the HST was pointed away from the radiant of the stream and in November, 2002 this direction
was nearly at the location of the Helix Nebula. This afforded the opportunity to make a series of 
observations during a 14 hour period. The program and its results (McCullough \etal\ 2002), which immediately went into the public domain,
are described on the web site http://archive.stsci.edu/hst/helix. In this study we use only the ACS WFC results,
these being a series of exposures in the F658N filter (transmitting equally well both the \Ha\ 6563 \AA\ and [N~II] 6583 \AA\ lines) and
the F502N filter (dominated by the [O~III] 5007 \AA\ line).  Nine overlapping fields were used centered at about 2\arcmin\
west of the central star and with the west boundary pointed to the north-northwest. The field covered is illustrated 
in \S\ 3.3 (Figure 7), where the images are first discussed.  The Helix Nebula is sufficiently large that even these nine fields could not cover the entire object,
only a significant portion of the main-ring of the object. Through an error, the central
star image fell into a gap between fields, but the field was otherwise continuous, with multiple exposures allowing
cosmic ray correction over most of the image. The total exposure times were t(F658N)=5.12 hours and t(F502N)=4.92 hours.
The individual field exposure times are much shorter than earlier WFPC2 observations (OH1996,OB1998) and since the instruments produce comparable signals
per pixel, the signal was much lower, although adequate for our purpose. The total of 46 exposures in each filter were
combined into a continuous mosaic by STScI staff members and we have used the data product where the signal from four adjacent pixels
was averaged, so that each pixel is a double sized 0.10\arcsec\ (comparable to the WFPC2 0.0996\arcsec\ pixels), but the images have a correspondingly higher signal to noise ratio.

\subsection{CTIO 4-m MOSAIC Imaging in Four Filters}
We have supplemented the HST ACS observations with lower spatial resolution images made with the MOSAIC camera
mounted at the prime focus of the Blanco 4-m telescope at CTIO. Because of equipment problems, we were able to
use data from only an array of four 2224x4096 pixel CCD detectors aligned north-south, instead of the usual square array
of eight detectors. Each detector had a gap of about 15\arcsec\ with its neighbor. The pixel size was 0.264\arcsec\ square
and the seeing varied from 0.9\arcsec\ to 1.1\arcsec\ FWHM (full width at half maximum) over the period of the observations (nights of 17 and 18 September,
2003). Four filters were used. The c6009 filter (FWHM=80 \AA, $\lambda \rm _{peak}$= 6563 \AA) isolated equally well the \Ha +[N~II] 
lines. The c6013 filter (FWHM=80 \AA, $\lambda \rm _{peak}$= 6725 \AA) isolated the red [S~II] doublet at 6716 \AA\ + 6731 \AA.
The c6014 filter (FWHM=50 \AA, $\lambda \rm _{peak}$= 4990 \AA) isolated the [O~III] line at 5007 \AA. Each of these were
146 mm square. A smaller (10 cm square) filter, kpno1567 (FWHM=25 \AA, $\lambda \rm _{peak}$= 4876 \AA) was also used for
isolation of the \Hb\ line at 4861 \AA. Each filter also passed a certain amount of continuum, which was always only a small
fraction of the total signal. Four images were made in each filter. The first pair covered the eastern part of the
nebula and the second pair an overlapping region of the western part of the nebula. The exposures in each pair were displaced 30\arcsec\ 
from one another. The exposure times were the same within each filter, being 300 seconds for c6009 and c6014,  900 seconds for c6013,
and 600 seconds for kpno1567. Flat field corrections were made using dome-flat exposures. A modification of the IRAF 
\footnote{IRAF is distributed by the National Optical
Astronomy Observatories, which is operated by the Association of
Universities for Research in Astronomy, Inc.\ under cooperative
agreement with the National Science Foundation.}
msc package of tasks was used for the data reduction. The images were calibrated into relative photon rate values using
the spectrophotometrically calibrated samples from O1998, using the assumption that each filter is dominated by
the target emission line.  An isolated [N~II] image was made by subtracting a scaled version of the \Hb\ image from
the \Ha + [N~II]. All of the images were put into the same reference frame, whose vertical axis had a position angle (PA)
of 0.9\arcdeg. Characteristic reproductions of 1500\arcsec x1653\arcsec portions of the images are shown in Figure 1.
For completeness, we have added the HeII  4686 \AA\ image of this same area from the 1997 Curtis-Schmidt observations
(O1998), which were made with 2.3\arcsec\ pixels and has the continuum and stars subtracted from it.

\subsection{Creation of  Combined CTIO and ACS Images}

\begin{figure}
\plotone{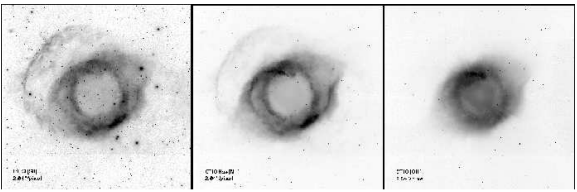}
\caption{
{This negative depiction of 1761\arcsec x 1761\arcsec\ CTIO fields centered on the central star of the Helix Nebula is
adjusted to bring out the outer parts of the nebula, as described in the text. The filling of the smaller ellipse 
is to indicate that this is actually the outer part of a filled disk, the innermost portion being visible primarily
in HeII emission.}
} \label{fig-5}
\end{figure}

Because of the almost identical transmission characteristics, we could combine the CTIO c6009 and ACS F658N images into
a single composite image isolating \Ha +[N~II] emission and the CTIO c6014 and ACS F502N images into a single composite image isolating
[O~III] emission. This was done by degrading the ACS pixels to those of the CTIO images (0.264\arcsec /pixel), aligning them
using the CTIO images as the reference, scaling the images to be the same in the outer regions of the ACS images,
then combining them so that the inner portion of the image comes from the HST-ACS and the outer from the CTIO MOSAIC.
In this way the inner portion has a higher resolution, being determined by the pixel size, while the 
outer portions have the resolution set by the seeing (about 1\arcsec) at the time of the CTIO observations. The  monochromatic \Ha +[N~II] 
and [O~III] images were then combined into a single color image, using red for \Ha +[N~II], green for an average of \Ha +[N~II] and [O~III], and blue for [O~III], the result 
being shown in Figure 2. This image is a good guide to the variation in the ionization structure of the nebula and the structure
within the nebula, as discussed extensively in the remainder of this paper.

\section{ANALYSIS}

This new data set from CTIO and HST images allows us to derive a more accurate picture of the structure
of the Helix Nebula. For the CTIO images the unprecedented diagnostic emission line isolation and high signal
to noise ratio is the primary advantage. In the case of the HST images, our advantage is in having 0.1\arcsec\
resolution over most of the region of the nebula that shows knots.

\subsection{Interpretation of the Main-Ring as a Pair of Rings}

We have used the CTIO images to reassess the validity of the widely accepted model (M1998, O1998, H1999)
that the main-ring emission of the nebula is explained as a thick disk, ionization bounded  along
its perimeter. Although the earlier imaging data used in modeling the nebula (O1998) easily 
resolved the object, our new observations give better coverage of different ionization stages and higher spatial
resolution. The primary motivation for reassessment was the impossibility of
explaining the velocity structure of the nebula by the simple disk model, which is discussed in \S\ 4.1.
However, the multiple loops within the main-ring, which give rise to the name Helix Nebula, 
have vexed researchers trying to model the nebula for decades. In fact, Taylor (1977) used the appearance
of the nebula and his few [O~I] velocities to construct a model of a cylindrical helical distribution of
material. Our improved emission line images and detailed velocity mapping allow us to produce a simpler model
than Taylor's helix, which produced an explanatory paper of very similar title to ours (Fabian \&\ Hansen 1979).
Our model is, however, more complex than the simple thick disk model.

Examination of our images in multiple emission lines indicates that there are two primary rings of
emission, each having ionization structure indicating that they are ionization bounded. In Figure 3
we show an overlap of two ellipses that fit an inner and an outer-ring within the main-ring of the
nebula. The inner-ring is an ellipse of major and minor axes 499\arcsec\ x 459\arcsec\ and the outer-ring is an 
ellipse of major and minor axes 742\arcsec\ x 446\arcsec. These correspond to circles of radius 250\arcsec\ and 371\arcsec\ 
(0.26 pc and 0.38 pc respectively) seen at tilt angles of 23\arcdeg\ and 53\arcdeg\ out of the plane of the sky.
The reality of two separate rings, inner-ring and outer-ring, is supported by three primary lines of evidence: 1) the visual 
appearance as outlined in Figure 3, 2) the evidence for two ionization fronts (see Figure 4 and discussion below), and 3) the 
distinct velocities of the two rings components (\S\ 4.1).
The discussion of nebular velocities in \S\ 4.1 indicate that the near side of the inner-ring is oriented
towards PA=288\arcdeg\ and the near side of the outer-ring is at PA=168\arcdeg. The former number may be
slightly small as we'll see in \S\ 4.3.2 that a feature associated with the inner-disk argues for PA=303\arcdeg.

The first evidence of two such rings lies in the imaging data of H1999, especially in their Figure 5, which
gives the [N~II]/\Ha\ ratio image, which indicates two zones of ionization stratification, although they
identified only one, associating it with the thick disk model. Our images in several stages of ionization 
provide similar but more extensive evidence. The expected photoionization stratification for a nebula
has been presented in simple form in O1998, which draws on the fact that the ultraviolet opacity to the stellar radiation
is dominated by hydrogen and helium, with the collisionally excited lines of the heavy ions usually providing the best tracers.
Closest to the very hot central star is the zone of 
H$^{+}$+He$^{++}$, which is traced only by the HeII emission, the strongest optical line of which is at 4686 \AA.
The next zone outward is the H$^{+}$+He$^{+}$, which is most easily traced by the strong [O~III] line at 5007  \AA.
The third zone outward is the H$^{+}$+He$\rm^{o}$, which is traced most easily by the strong [N~II] doublet lines at
6583 \AA\ and 6548 \AA. Next is the hydrogen ionization front, where the best tracer is the [S~II] doublet lines
at 6716 \AA\ and 6731 \AA. 

We have looked for evidence of this ionization stratification by examining a sample of our monochromatic CTIO images.
The sample was 57.5\arcsec\ wide and extended across the center of the nebula in an east-west direction. The
stars within the sample were edited out (local surface brightness values being substituted). The 2-D array was 
reduced to averages using IRAF task {\it splot} and the results are presented in Figure 4.
The signal from each ion was scaled for clarity. The lower spatial resolution of the HeII data used (2.3\arcsec , O1998)
should make no difference since the HeII peak shows no evidence of fine-scale structure.

We see in Figure 4 that the H$^{+}$+He$^{++}$ zone is limited only to the center near the central star, which
indicates (O1998) that the nebula contains material up to very close to the central star. The ring (rather than a disk) appearance of the nebula is
caused by the fact that HeII emission is weak and not usually included in images of the nebula. Immediately 
outside of this central HeII peak we see a transition to the H$^{+}$+He$^{+}$ zone as traced by [O~III], followed
by the H$^{+}$+He$\rm^{o}$ zone traced by [N~II], and finally an ionization front traced by [S~II]. The close overlap
of the profile of [S~II] and [N~II] emission indicates the important role of charge exchange at the low densities that
apply in the nebula (O1998), which broadens the [S~II] emitting zone and the fact that the zones must be curved and the
observer looks along a cross-section through the nebula. These first features are all associated with the inner-ring,
which is actually a disk with a HeII core. The inner-ring structure seen in Figure 3 is interpreted as the outer, lower
ionization portions of the disk that forms the inner structure of the nebula, which we will call the inner-disk.

In Figure 4 we see a small rise in the [O~III] surface brightness on both sides of the central star at distances
of about 110\arcsec. This was first noted by Meaburn \&\ White (1982).  The fact that this peak of emission 
occurs without an accompanying lower ionization peak outside of it indicates that it is a peak in density, rather than
a density peak with an associated ionization front. This density peak is probably associated with the inner H$^{+}$+He$^{++}$
zone as it occurs just outside of where the HeII emission begins to drop in surface brightness.

There is a second progression of ionization structures further out, which
correspond to the outer-ring structure. It has no obvious HeII core, nor would one be expected at these greater 
distances, [O~III] is weaker than in the inner-ring, and the [N~II] and [S~II] features are relatively stronger than
on the inside. Beyond the outer-ring one sees [S~II] features that result from the outer-nebula. In \S\ 4.5 
we address the more difficult problem of the structure of the rings out of their planes.

\begin{figure}
\plotone{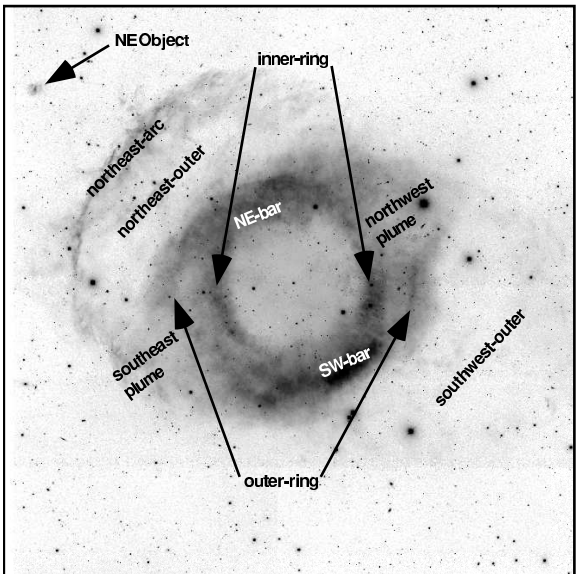}
\caption{
{This depiction of the the [S~II] image from Figure 5 has superimposed labels that indicate features 
discussed in the text. The inner-ring and outer-ring features together compose the main-ring, where the bulk of 
the optical emission from the nebula arises. The inner-ring is the low ionization portion of the inner-disk.}
} \label{fig-6}
\end{figure}

\subsection{Structure of the Outer-Nebula}

Outside of the main-ring of the nebula is the structure we call the outer-nebula. This was first seen in 
photographic images (Malin 1982), the velocity was measured in one eastern region (Walsh \&\ Meaburn 1987), the structure
was more clearly delineated at 2.3\arcsec\ resolution in O1998, and fainter extensions were seen in low resolution
(48\arcsec\ pixels) \Ha\ images (Gaustad \etal\ 2001, S2002). Our large filter CTIO images allow delineation of this structure at
an unprecedented combination of spatial and  spectral resolutions and reaches to surface brightness levels comparable to
those of the low spatial resolution study (S2002). A homogeneous sampling of our results are shown in Figure 5, where we depict in
negative our three CTIO images over an 1761\arcsec x 1761\arcsec\ field. This is similar to Figure 1 in that the 
images saturate at single regions in the inner ring, but are now displayed with a contrast value of $\gamma$=1.5, which
enhances the appearance of the lowest brightness levels, and the FOV is larger. In addition, we have not attempted to separate the \Ha\ and
[N~II] components because our reference \Hb\ image was made with a smaller filter that did not give homogeneous 
illumination in the outer-nebula region and it is not safe to assume that the \Ha /\Hb\ ratio is the same there as it is
in the main-ring regions that were used for spectrophotometric calibration. If the [O~III] image is examined at its
faintest levels, one can see an outer structure, but this signal is probably due to nebular continuum combined with
scattered light, rather then being the [O~III] emission line.

In these images we see a well defined north-east arc, which shows much more fine structure in [S~II] than in the
\Ha +[N~II] image.  This structure is clearest to the north-east, but can be traced at smaller distances until it
joins the outer parts of the main-ring to the north northwest and the south.  We call this entire structure the
northeast-outer feature.
One sees in [S~II] a fainter emission 
beyond the north-east arc extending towards PA$\simeq$ 80\arcdeg. There is also a region of extended emission out to
the edge of the image with a symmetry axis of PA$\simeq$ 250\arcdeg, which we call the southwest outer feature.
For clarity, Figure 6 gives a set of labels for the various features  already mentioned, together with some 
introduced later. 


\subsection{Analysis of the Knots}

In the O2002 analysis of four nearby PN (including the Helix Nebula) it was argued that knots form
as instabilities at the ionization front and then are sculpted following passage of the front.
The more complete mapping of the inner-disk by the ACS allows us to evaluate and refine that argument.
This evaluation was first done morphologically and then by quantitative analysis of these images.

\subsubsection{The Progression of Appearance of the Knots}
We examined the entire ACS mosaic in detail, looking for patterns in the appearance of the knots. 
Since space does not allow us to reproduce the entire image at full spatial resolution, we selected
four sample areas to illustrate the pattern that was found. The regions sampled are shown in Figure 7.
The first four are all associated with the inner-disk and sample five is associated with the outer-ring.
Higher resolution images of these sampled regions are shown in Figures 8-9-10.

\begin{figure}
\plotone{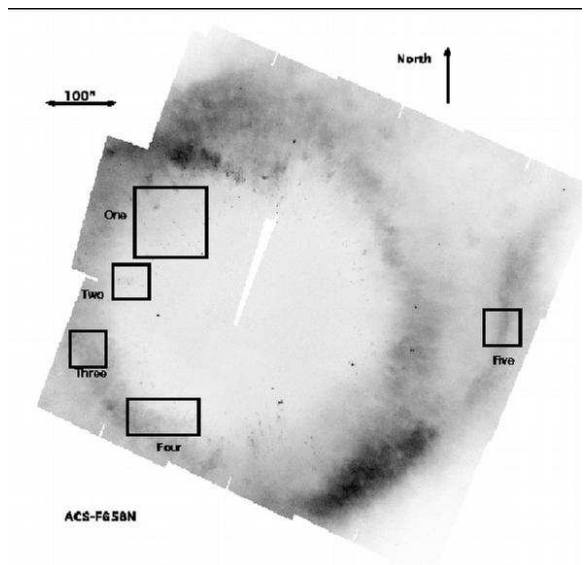}
\caption{
{This reduced resolution, negative depiction of the ACS \Ha +[N~II] mosaic image has superimposed the fields 
shown in the next three figures in both \Ha +[N~II] and [O~III]. Sample One is 102\arcsec x 102\arcsec,
Samples Two, Three, and Five are 51\arcsec x 51\arcsec. Sample Four is 51\arcsec x 102\arcsec.}
} \label{fig-7}
\end{figure}

\begin{figure}
\plotone{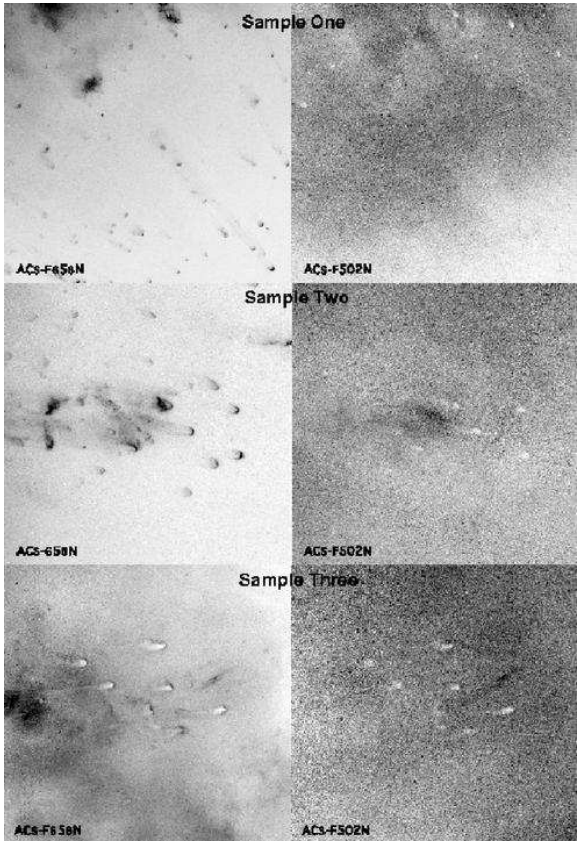}
\caption{
{This series of negative images shows the first three samples of Figure 7. Note that the scale of
the top sample (One) is twice that of the other two.}
} \label{fig-8}
\end{figure}

\begin{figure}
\plotone{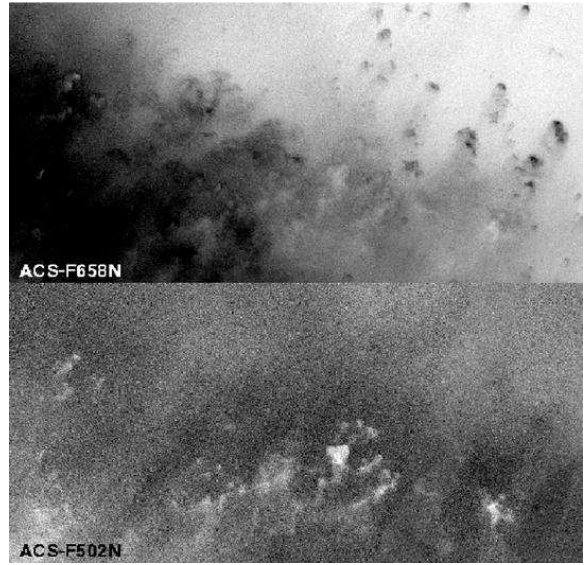}
\caption{
{This negative depiction of Sample Four from Figure 7 shows the properties of knots closest to the
nebular ionization front within the inner-disk.}
} \label{fig-9}
\end{figure}

\begin{figure}
\plotone{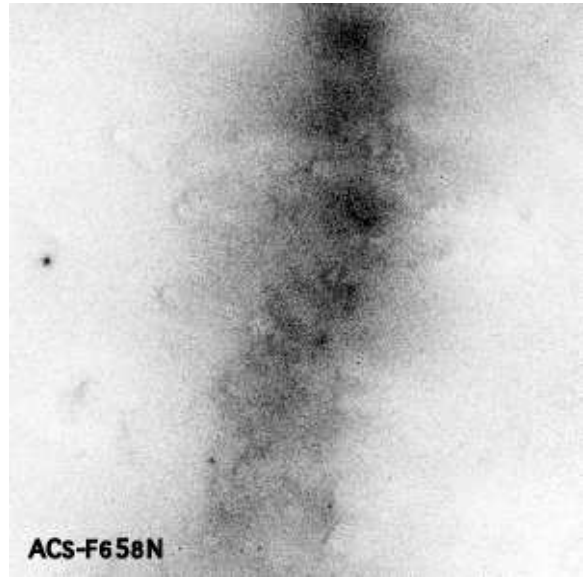}
\caption{
{This negative depiction of Sample Five shows that sculpted knots are found ahead of the nebular ionization
front in the outer-ring of the nebula. Only the \Ha +[N~II] image is shown since the [O~III] image is featureless.}
} \label{fig-10}
\end{figure}

We found that a clear pattern of forms of the knots was obvious. When the knots are well ahead of the
present location of the nebula's ionization front they are well sculpted. The knots have a well-defined
ionized cusp on the central star side of a crisp teardrop shaped extinction core, the latter best seen in the [O~III]
images. These objects are illustrated in Figure 8, Sample One. At distances slightly closer to the nebular 
ionization front we begin to see that the extinction cores are more diffuse, as shown in Figure 8, Sample
Two. Closer yet to the nebular ionization front we see that the extinction cores are even more diffuse and we
begin to see them in extinction in the \Ha +[N~II] images of the knots, as shown in Sample Three of Figure 
8. This pattern is continued even closer to the nebular ionization front, as illustrated in Figure 9,
where we see that the bright cusps are much more irregular and the boundaries of their tails are not simple
shadow projections lying along radial lines drawn toward the central star. The cores and tails are dark in the
\Ha +[N~II] images and very dark in the [O~III] images.

We interpret this progression to mean that the knots are concentrations of dust and gas that form in or
immediately outside of the nebular ionization front. The ionization front moves outward as the nebula expands
and attains a lower average density, leaving the knots behind (although the knots are still expanding away
from central star). The knots have enough material to be optically thick to Lyman Continuum (LyC) radiation, thus forming
a small local ionization front and initiating the process of photo-ablation of material through the over-pressure
condition that results from photoionization. From the start there is more material towards the center and 
photo-ablation soon has removed the lower density material outside of the shadow-cone (Cant\'o \etal\ 1998) cast by the core 
of the knot, whereas the shadowed material is lost more slowly.  

We see the same pattern being  repeated in the outer-ring as shown in Figure 10, although this figure
samples but a single area of that structure.

\subsubsection{Quantitative Analysis of the ACS Images of the Knots}

\begin{figure}
\plotone{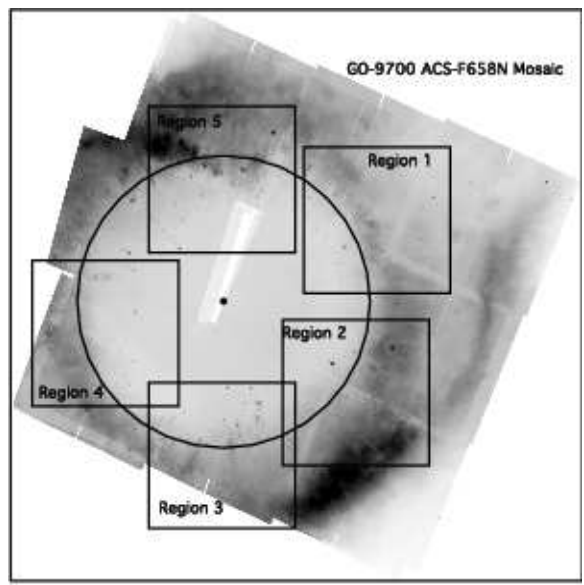}
\caption{
{This negative depiction of the ACS \Ha +[N~II] mosaic has superimposed the boundaries of the regions
for which a quantitative assessment of the properties of the knots was conducted. The circle indicates
a distance of 200\arcsec\ from  the central star, which is shown as a filled circle.}
} \label{fig-11}
\end{figure}

We also looked for a more quantitative means of assessing the properties of the knots. 
We sampled five regions of the ACS \Ha +[N~II] mosaic, as shown in Figure 11. Within these samples
Region 1 and Region 2 offer the chance to examine knots whose distances from the central star can be used as 
measures of the distance from the nebular ionization front for either the inner-disk or the outer-ring.
A general property of the knots is that the surface brightness at the substellar apex of their ionized cusp
decreases with distance from the central star (OH1996) which to the first order can be explained by there
being a decreasing flux of LyC photons at greater distances, and the deviations from this simple theory 
can be explained by loss of LyC photons prior to reaching the ionization front (L\'opez-Mart\'{\i}n \etal\ 2001).
However, our analysis of the forms of the
knots indicated that knots closer to their associated nebular ionization front were darker in the middle,
reflecting the fact that their cores had suffered little mass-loss. 

\begin{figure}
\plotone{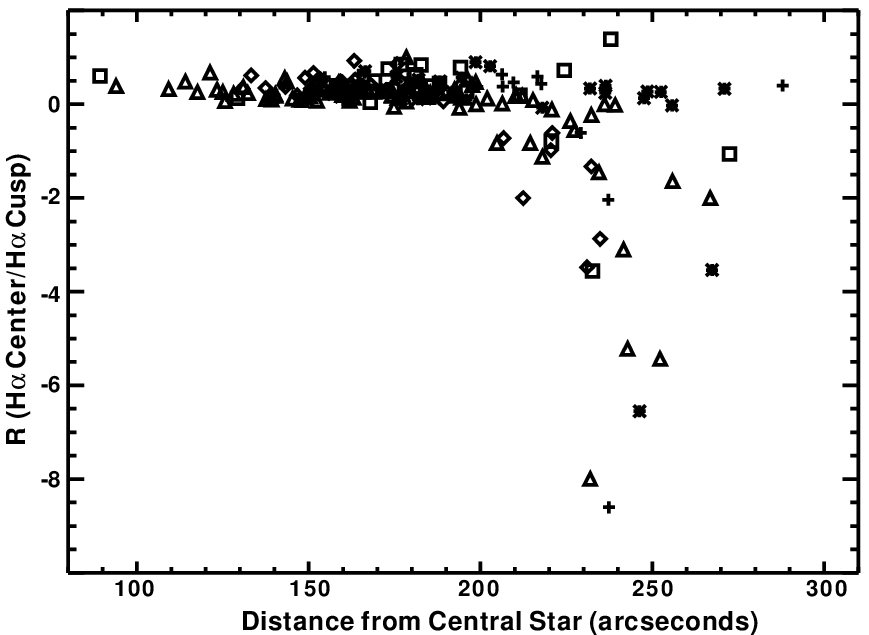}
\caption{
{This figure plots the ratio of the surface brightness of the core of the knots to the surface brightness 
of the apex of the bright cusp (both in \Ha +[N~II]) as a function of angular distance from the central 
star.  The different regions are coded by symbols: 1(+), 2(*), 3(open rectangles), 4(diamonds), 5(triangles).}
} \label{fig-12}
\end{figure}

We have examined the variation in the amount of extinction occurring in the core of the knots by sampling
the cusp in an 0.3\arcsec\ square centered on the apex of the bright cusp and on the center of the core. 
When the position of the latter was not obvious, it was determined from the shape of the well-defined cusp.
In each case two adjacent regions of the nebula were measured to determine the background, which was then
subtracted from the signal. A negative signal occurs if the core has a lower signal than the ambient nebular emission.
A ratio (R) of the center of the core signal to that of the cusp was calculated and
the results are presented in Figure 12.  

We see from Figure 12 that R is a small positive number for most of the knots in the five
regions sampled. However, as one goes beyond 200\arcsec\ from the central star that some of
the knots have negative values of R, indicating that the cores are dark in \Ha +[N~II]. When we examine the
figure in detail we see that the expectations of \S\ 3.3.2 are realized. We begin to see a wide range of
values of R at large distances because some of these knots are close to their associated nebular ionization
fronts (hence have large amounts of core extinction) while others seen at the same distance are further
removed from their nebular ionization fronts. This is because some are associated with the inner-disk and others the 
outer-ring.

\subsection{The Radial-Rays on the Outside of the Main-Ring of the Nebula}

\begin{figure}
\plotone{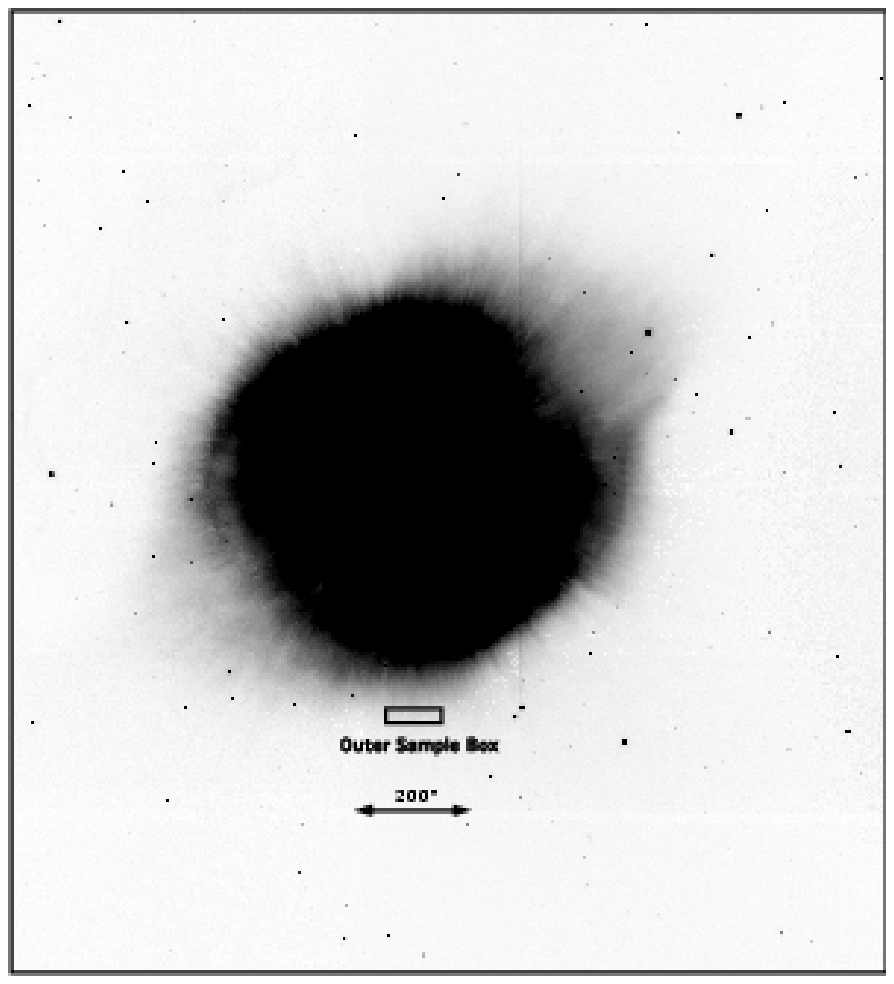}
\caption{
{This negative depiction of the CTIO [O~III] image has had the saturation level set low enough that
one can see detail in the region immediately outside the main-ring. We see about 108 radial-rays along
the circumference of the nebula. The region marked Outer Sample Box and its use are described in the text.}
} \label{fig-13}
\end{figure}

\begin{figure}
\plotone{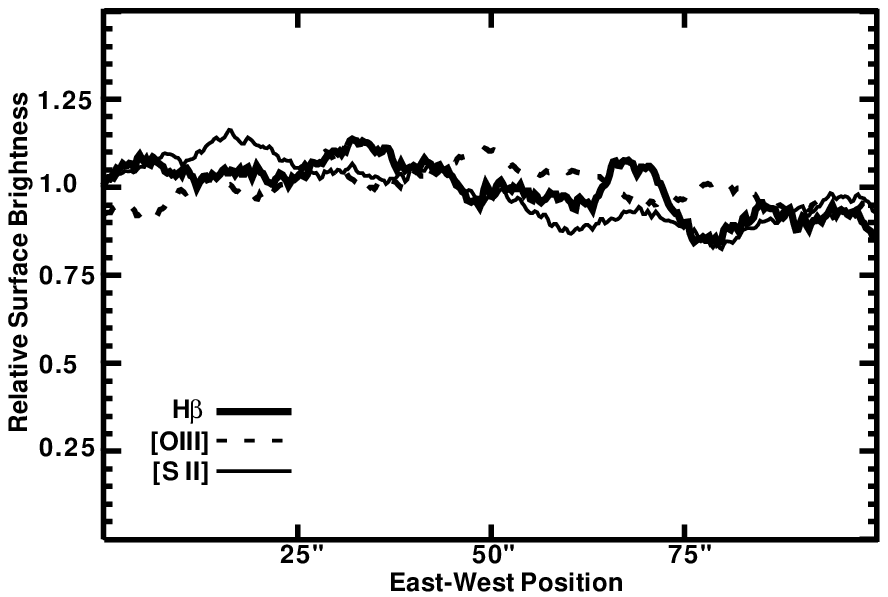}
\caption{
{The intensity within the region called the Outer Sample Box in Figure 13 is shown for \Hb , [O~III],
and [S~II] emission. The signal from each line has been normalized to unity and the values averaged along
the shorter, north-south, dimension.}
} \label{fig-14}
\end{figure}

The region immediately outside of the main-ring shows a peculiar radial pattern of light and dark, which
we call the radial-rays. These were first visible in the ratio of [O~III] to \Hb\ image of O1998 (his Figure 1).
Within the main-ring H1999 found similar structure in their images of the [N~II] to \Ha\ ratios, where they
were called plumes-a term used differently in the present paper (\S\ 1, Figure 6). The main-ring radial features are
extensions of the ionization shadow-cones of the optically thick knots, as suggested by Cant\'o \etal\ (1998)
and discussed in detail in O'Dell (2000, henceforth O2000). 

In order to quantitatively characterize the radial-rays, we have sampled a 96\arcsec x 24\arcsec\ region
centered 412\arcsec\ due south of the central star, as shown in Figure 13. This was done in the
CTIO \Hb , [S~II], and [O~III] images because only they cover the outer region at good resolution. These three filter
images give good isolation of emission, whereas the other CTIO images require an assumption of a constant
\Ha /\Hb\ ratio for deriving the [N~II] emission from the \Ha +[N~II] image. At this distance and with this
narrow of a sample the effect of the divergence of the radial-rays is small and we have traced the profile of
the average image along an east-west direction, as shown in Figure 14, where the signals in each emission-line
have been normalized to unity. We see that there are periodic fluctuations in both lines with amplitudes of about
$\pm$6, 5, and 3\%\ in intensity in \Hb , [O~III], and [S~II] respectively and scale widths of about 8\arcsec, there being eight \Hb\ peaks in the sample. It is
significant that the intensity of the \Hb\ and [O~III] signal is anti-correlated, i.e. as one goes up, the 
other goes down. The pattern of variation of the [S~II] emission is harder to define, but it appears to increase
with increases of the \Hb\ emission.
If this sample is characteristic of the entire nebula, then there would be about 108 
radial-rays around the full circumference. Their nature and interpretation are discussed in \S\ 4.2.

\section{DISCUSSION}

The rich variety of images of the Helix Nebula in various emission lines (both optical and radio)
and the existence of a useful amount of high velocity resolution spectra has allowed us to develop a
refined model for the object. In this section we present a discussion of the properties and construction
of the various parts of the nebula, then present the integrated picture.

\subsection{The Dynamics of the two components of the Main-Ring}

In \S\ 3.1 we produced evidence that the 
the eponymous helical structure in the plane of the sky is the result of two circular rings
of emission having different orientations. This model is confirmed through examination of
the available radial velocity data. Most of this velocity data is associated with the knots.

There is reason to argue that all of the measured CO sources seen in the Y1999
study are associated with knots. As the spatial resolution of the CO observations
have improved ( Huggins \&\ Healy 1986, Healy \&\ Huggins 1990, Forveille \&\
Huggins 1991, Huggins \etal\ 1992, Y1999),
there has been a progressive ability to isolate individual
knots. The best resolution study (H2002) targeted an individual knot (378-801).
Even at the lower resolution of the Y1999 study, spectra of individual
data samples commonly show multiple peaks of emission at various nearby velocities,
indicating that the unresolved regions are actually composed
of multiple emitters. A similar progression of improved resolution in \htwo\
observations has produced clear evidence for association with knots (S2002).
The isolation of individual knots becomes more difficult as one goes farther from the
central star because the number surface density of knots increases rapidly and 
the lower resolution CO studies appear amorphous first, and the higher resolution \htwo\
studies only appear amorphous in the outermost regions. 

Y1999 point out that dynamically there appear to be two types of CO emitters. The first type
is a group of small sources all found within 300\arcsec\ of the central star. The velocities
of these sources are distributed as if they all belong to an expanding torus region, which 
produces a nearly sinusoidal variation in the radial velocities with an amplitude of $\pm$17 
\kms , as found earlier by Healy \&\ Huggins (1990). They call these sources the inner-ring.
The second type of emitters are found in the regions they call the outer-arcs. The western
outer-arc starts to the north of the central star and spirals out before returning to a closer
distance to the south of the central star. The east outer-arc starts to the south-south-west from
the central star, reaches its greatest distance and surface brightness west of the central
star, then becomes closer and reaches the main emitting ring of the nebula to the north of the
central star.  The west outer-arc is clearly interrupted to the northwest, where the 3-D
model of the nebula (O1998) argues is the location of the extended symmetry axis of the bipolar model.
We show in Figure 15 the Y1999 CO data combined with the results from Taylor (1977),
who measured velocities in [O~I] at several points in the main body of the nebula.
The coincidence of appearance and the kinematic structure developed in this section establish
that these CO outer-arcs are components of the outer-ring introduced in \S\ 3.1. 

Do the outer-arcs contain a series of discrete knots, like the inner-disk? This is very likely since
the higher spatial resolution \htwo\ images of S2002 show a multitude of small sources and 
the high velocity resolution CO study of Y1999 shows multiple emitters.  The excellent resolution
\Ha +[N~II] images of M1998 do not extend into the regions of the outer-arcs. Our CTIO images
of slightly lower resolution than M1998's cover the outer-arcs and show them to break down
into individual knots of bright and dark (Figure 10).

\begin{figure}
\plotone{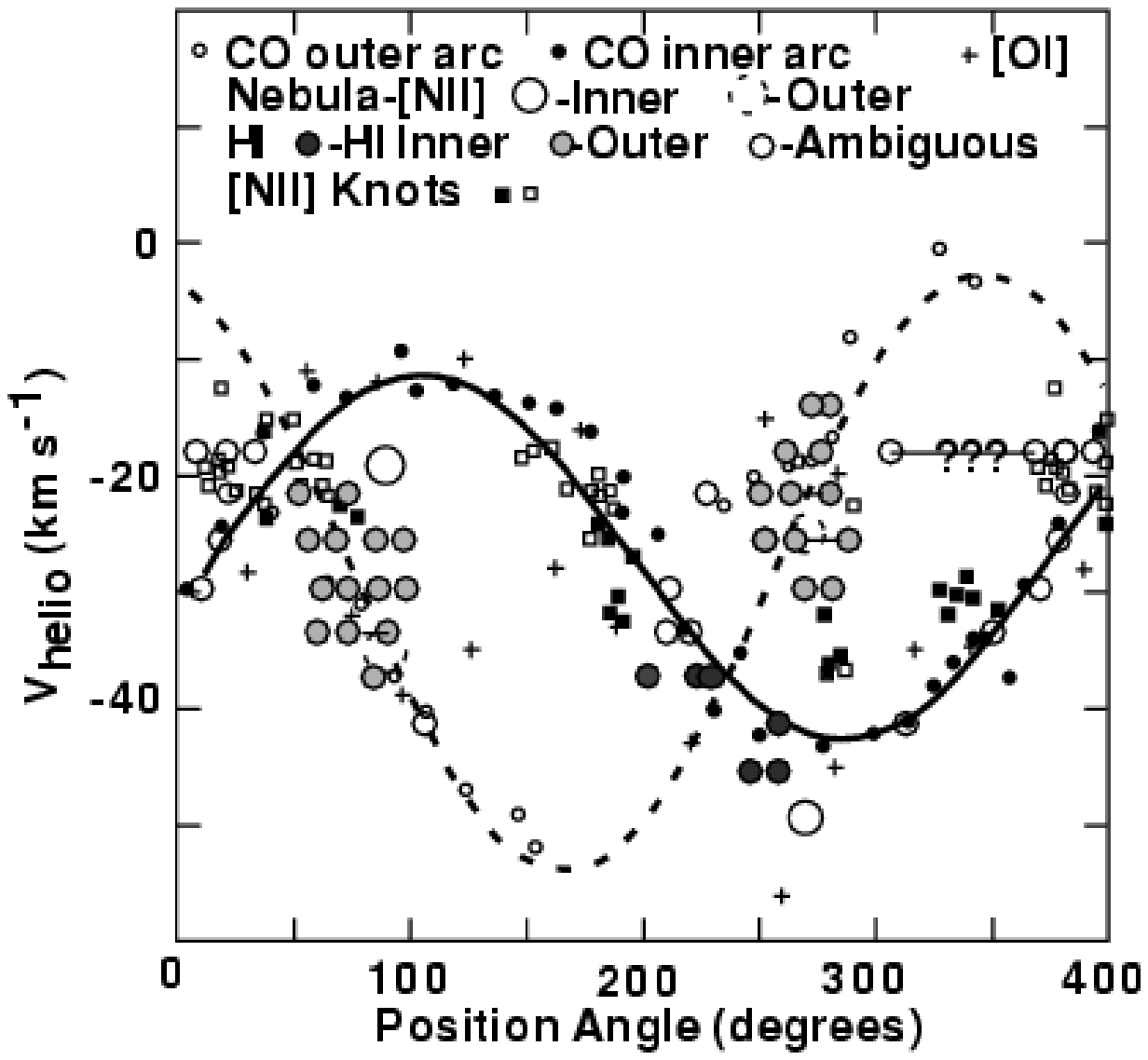}
\caption{
{We have plotted here the heliocentric radial velocities for various discrete sources in the Helix
Nebula as a function of their position angle from the central star. The coding of many of the 
symbols are given across the top. The squares indicate 
[N~II] velocities given in
M1998 for knots that have low (open) or high (filled) amounts of extinction in [O~III] images. The CO
velocities for the inner and outer-arcs are from Y1999 and the [O~I] velocities are from 
Taylor (1977). The Nebula-[N~II] velocities are from M1998 and the H~I velocities are from the
21-cm observations of Rodr\'{\i}guez, Goss, \&\ Williams (2002). The superimposed sine waves have
amplitudes of $\pm$25.5 \kms\ (outer-ring, dashed line) and $\pm$15.5 \kms\ (inner-ring, solid line) and peak negative velocities
at 168\arcdeg\ (outer-ring) and 288\arcdeg\ (inner-ring). The H~I feature with question marks indicates a feature
with no corresponding optical component.}
} \label{fig-15}
\end{figure}

\begin{figure}
\plotone{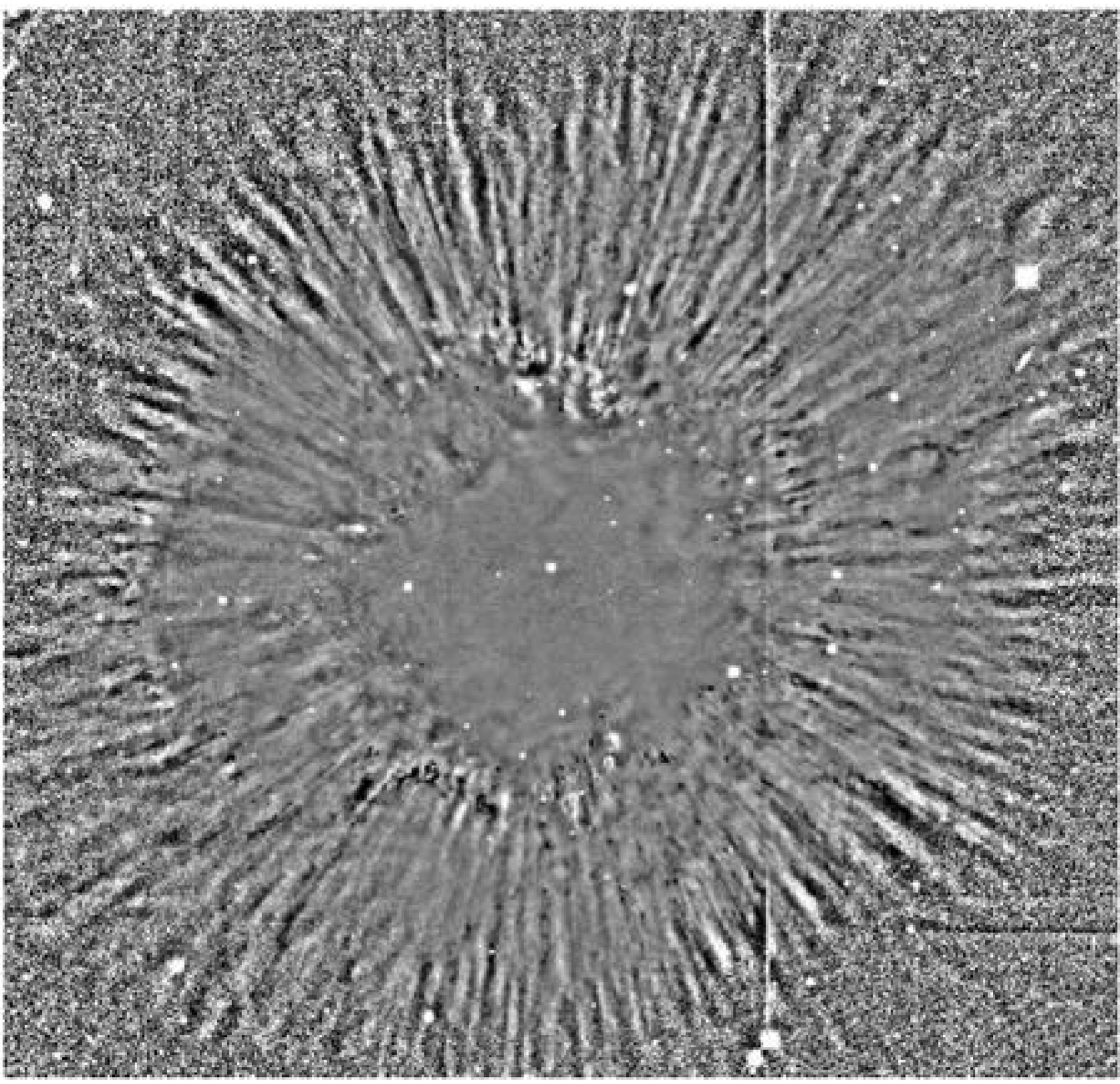}
\caption{
{In this [O~III] image of the central 944\arcsec x905\arcsec\ of the nebula we have enhanced the visibility of the 
small features by taking the ratio of the original image to its local average value as described in \S\ 4.2. One 
can see the individual inner knots as crisp dark features and the disconnected radial-rays extending out beyond
what is usually considered the limit of the main-ring of the nebula.}
} \label{fig-16}
\end{figure}

The Helix Nebula has been mapped by  Rodr\'{\i}quez, Goss, \&\ Williams (2002) in the H~I 21-cm line with 
a beam of 42\arcsec x 44\arcsec.
This spatial resolution is sufficient to allow identification of the source
with the inner or outer-rings of emission that we see in Figure 3 and these results have been
added to Figure 15. We see that interpretation of the H~I emission as coming from the
same areas as the CO presents a self-consistent result, i.e., that they too come from two velocity
systems. This emission is almost certainly from the knots, as the CO observations clearly
isolate individual knots, including ones that obviously fall well within the ionized part of
the nebula and removed from any photon dominated region that lies outside the ionization front of the nebula itself.
Rodr\'{\i}quez, Goss, \&\ Williams (2002) argue that there is a slight displacement of the H~I emission outside of the zone of ionized
gas, but this evidence is weak and probably an artifact of their having taken a profile averaged 
over all position angles, a dangerous procedure because of the asymmetry of the nebula.

M1998 present radial velocities (in [N~II]) for 50 knots over a range of position angles. We have 
added those results to Figure 15 and see that the velocity patterns agree quite well with the
results from the CO, H~I, and [O~I]. M1998 argues, as was done in their earlier paper (Meaburn \etal\ 1996),
that their knot velocities indicate that the knots are expanding at only about one half of the
expansion velocity of the nebula. This conclusion is based on a comparison of the radial velocities
of the system of knots with the expansion velocity as measured along a line of sight through
the central star.
We test this interpretation with the nebula expansion data from M1998 in the [N~II] line. Specifically, from their Figure 5.A, we extract
the radial velocities of the nebula in the inner and outer-rings
at position angles of 90\arcdeg\ and 270\arcdeg, at which points the two rings are separated unambiguously both on the sky and in velocity. Including these data in Figure 15 shows that the nebular material is flowing at essentially the same velocity of the co-located knots.

The resulting kinematic picture is entirely consistent with the disk and ring model derived from the
appearance of the nebula. There are two velocity systems, the easily traced inner-ring portion of the inner-disk which expands with a maximum
component of 15.5 \kms\ and an outer-ring expanding with a maximum component of 25.5 \kms.
When one takes into consideration the tilts of the rotation axis of the rings (23\arcdeg\ for
the inner-ring and 53\arcdeg\ for the outer-ring, the spatial expansion velocities become
40 \kms\ (inner) and 32 \kms\ (outer). If these expansion velocities have been maintained since
early in the history of the two rings, the corresponding ages are 6560 yrs (inner) and 12,100 years (outer).

There is no published velocity information for the HeII emission region that lies in the core of
the nebula. However, it would be very important to obtain this information because it should help
to explain how material can continue to exist in this region. The kinematics of the HeII may also
lead to an explanation of the presence of the 110" radius rise in [O~III] emission that occurs within the inner-disk (\S\ 3.1) just
outside of the HeII core.

The inner-ring is the low ionization portion of an inner-disk, with the HeII emitting zone at its center.
Within this disk are two unresolved x-ray sources (Leahy, Zhang, and Kwok 1994, Leahy et~al. 1996, Guerrero et~al. 2001),
the lower temperature component being continuum emission from the central star.
The central star has no detectable stellar wind (Cerruti-Sola 
\&\ Perinotto 1985) so that the million degree gas in the hotter source is unlikely to arise from a stellar wind or the shocks that it can cause,
a conclusion strengthened (Chu et~al. 2004) by there not being O~VI emission or absorption.
Guerrero et~al. (2001) argue that the hotter source is probably an M-dwarf companion to the central star.
Examination of the original CTIO HeII images shows no evidence down to about 5\arcsec\ from the star for the dip in HeII that would result if
there was an extended hot core, which would have a low emissivity in this HeII recombination line.
The fact that the inner-disk has no shocked region is not in accord with the two-wind model of the development
of the PN, unless the original wind-blown cavity has been refilled by material from the star or
backflowing from an earlier shell structure. Certainly the central star is now so far evolved that it does not
have a stellar wind, but the absence of a shocked region would demand that it has cooled below the level of detection
since the time that the wind stopped. 
There is no dynamical evidence of a current backflow of nebular material . In fact, the limited velocity data indicate that the nebula
is expanding homologously as do most other PN. 

How such a double structure can be formed is problematic. Rotationally symmetric figures are common among the
PN. If the two structures indicate significant mass-loss at different times, then this would require a rapid
change of the axis (86\arcdeg ) in only 5500 years. As we'll show in several of the following sections, there
certainly is evidence for other periods of mass-loss with orientations different from either component of the
main-ring, which means that there is ample evidence for a wandering axis of orientation.

\subsection{The Radial-Rays surrounding the Main-Ring}

The radial-rays that we see outside of the main-ring contain considerable information about the structure
of the inner nebula. The basic theory that describes ionization shadows behind objects like the knots is
presented in Cant\'o \etal\ 1998 and the successful application of that theory to the Helix Nebula is
presented in O2000. Within a shadowed region illumination is by scattered LyC photons. 
These LyC photons are generated by recombination of ionized hydrogen selectively occurring close to the Lyman limit, whereas
the star's emission occurs at higher energies. This means that the shadowed region is illuminated by photons with a 
significantly smaller energy excess above the ionization threshold. As a result, the electron temperature 
in the shadow is about half that of the ambient nebula. This enhances the relative emission of a recombination
line, such as \Hb, and decreases the emissivity of collisionally excited lines, like [O~III]. This is
why one sees an anti-correlation of the brightness of the tails outside of the knots (O2000) and we see
in this study that this pattern continues into the outer regions, where we see the radial-rays.  

There is an additional process that contributes to the diminution of [O~III] in the tails of
knots, which is not covered in the Cant\'o et~al. (1998) paper, but draws on the ionization structure
of the nebula that is reviewed in \S\ 3.1.
The photoionization of the shadowed
regions is determined by the recombination continuum of not only H$^{+}$, but also He$^{+}$ and He$^{++}$.
In the innermost zone, where  He$^{++}$ is found and the nebula is traced by the HeII 4686 \AA\ recombination line,
the shadowed region sees a continuum at and just above 54.4 eV, which is sufficient to create  O$^{+3}$ because the
ionization energy for O$^{++}$ is 54.9 eV. This means that the nebular material and the shadowed regions have
the same preferred state of ionization of oxygen.  The same is true of the H$^{+}$+He$\rm ^{o}$ zone, where the shadowed
region is illuminated by the recombining  H$^{+}$ at just above 13.6 eV, which is enough to photoionize neutral 
oxygen into O$^{+}$.  In the H$^{+}$+He$^{+}$ zone O$^{+2}$ is dominant in the nebula because the
stellar continuum of less than 54.4 eV is adequate to photoionize oxygen from O$^{+}$ to O$^{+2}$ by the absorption
of photons of greater than 35.1 eV. However, the shadowed regions are of lower ionization. This is because in this zone
the shadows are illuminated by recombining He$^{+}$ at just above 24.6 eV and few of these will have enough energy
to photoionize oxygen, leaving it as O$^{+}$. This means that [O~III] cannot be produced within the shadows in
this region, which is where most of the knots and tails are found, but, they should be enhanced in [O~II] emission.

The large-scale structure of the radial-rays are illustrated in Figure 16, which is an [O~III] image that has
been divided by a median smoothed version of that same image using a square kernel 85 pixels (22\arcsec)wide. This
represents the ratio of a given pixel to the local average and allows one to easily see small local enhancements
and deficits.  

The weak positive correlation of \Hb\ and [S~II] is more difficult to explain. On the one hand [S~II] emission
demands collisional excitation of the ion's outermost bound electron, hence the emissivity per unit ion and
electron density increases with electron temperature. On the other hand, neutral sulphur has a lower ionizaton
energy than hydrogen, so that S~II should exist in a neutral hydrogen zone, disappearing quickly through
photoionization into S~III within an ionized hydrogen zone (O1998). This means that normally one expects to see
[S~II] emission only from the immediate vicinity of an ionization front, where one has both some population of
S$^{+}$ and electrons to do the collisional excitation.  This simplistic picture is blurred by the fact that
charge-exchange processes will widen the width where this emission can occur. O2000 presented arguments that
the tails immediately behind the knots in the Helix Nebula were fully ionized. 

When we examined the tails of the inner knots in the CTIO images one sees the anti-correlations seen in O1998 in \Hb\ and
[O~III]. However, he did not have [S~II] images and in our CTIO images we see that even in the tails of the 
inner knots that [S~II] is enhanced along with \Hb, indicating that the same process is occurring there as in
the outer radial-rays.  Most likely this is a charge-exchange dominated process that fills the otherwise
fully ionized tail regions.

The fact that there is some emission from the nebula, including [O~III], beyond the ionization boundaries 
identified to exist for the inner-disk and the outer-ring means that the nebula is not optically thick to LyC
at those projected distances 
above and below the planes of these rings, i.e. at high elevation angles.  The ``hocky-puck'' shape of the inner-disk has
a lower density distribution of gas above it that extends out to angular distances beyond the main-ring. Because there is much less material along a line of sight at
higher elevation angles the ionization fronts would be found much further away from the central star. In fact, there
may be no ionization front at very high angles. This means that the radial-rays are ionization shadows being formed
above the projection of the thick disk, 
probably at about 30\arcdeg\ above and below the plane of the inner-disk. This conclusion is strengthened by the fact that there are only about 108 of them,
whereas there are about 3500 knots (OH1996) in all the nebula.  This means that the radial-rays are formed by only the knots
at sufficiently high angles that their shadows extend beyond the ordinary ionization fronts of the inner and outer-rings.
When we examine the combined ACS+CTIO mosaic and Figure 16, we see no alignment of the radial-rays with any of the
knots seen in the inner parts of the nebula, rather, the associated knots fall further out, within the main-ring
structure of the nebula. These features are not individual knots, rather, groupings of them. This is consistent with the fact that outer-rays are much larger than the projection of single knots located in the inner parts of the nebula.

\subsection{The Structure of the Outer Nebula}

There are multiple structures that lie outside of the main-ring of the nebula. There are two (the plumes) which appear to
be contiguous with the inner-ring structure while the rest have no obvious link.

\subsubsection{The outermost structures}

The appearance of the outer nebula has developed gradually. It was initially reported by Malin (1982) from
broadband filter photographic images and in many ways his illustration remains the best until our Figure 17,
which covers an almost identical FOV. There are three primary features in these images.

\begin{figure}
\plotone{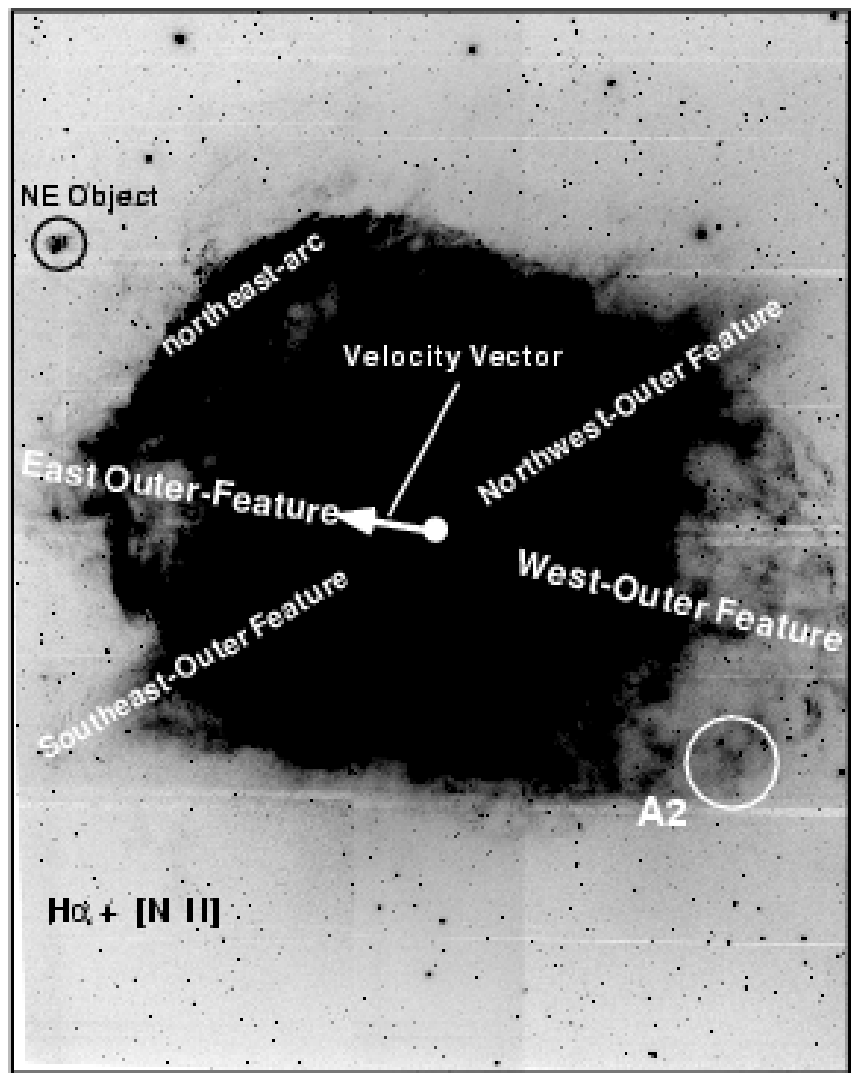}
\caption{
{This negative depiction of the full FOV (1759\arcsec x2229\arcsec) in \Ha +[N~II] shows only the small
surface brightness range from near the sky level to the outer parts of the main-ring and the northeast-arc. 
The location of the central star is indicated by the filled white circle and other features discussed in the text
in \S\ 4.3 are labeled, as is the direction of the adopted proper motion of the central star.
Imperfections in the background sky subtraction necessary in making a mosaic image from multiple detectors and
settings produce the narrow contrasting bands and the vertical linear features.}
} \label{fig-17}
\end{figure}

\begin{figure}
\plotone{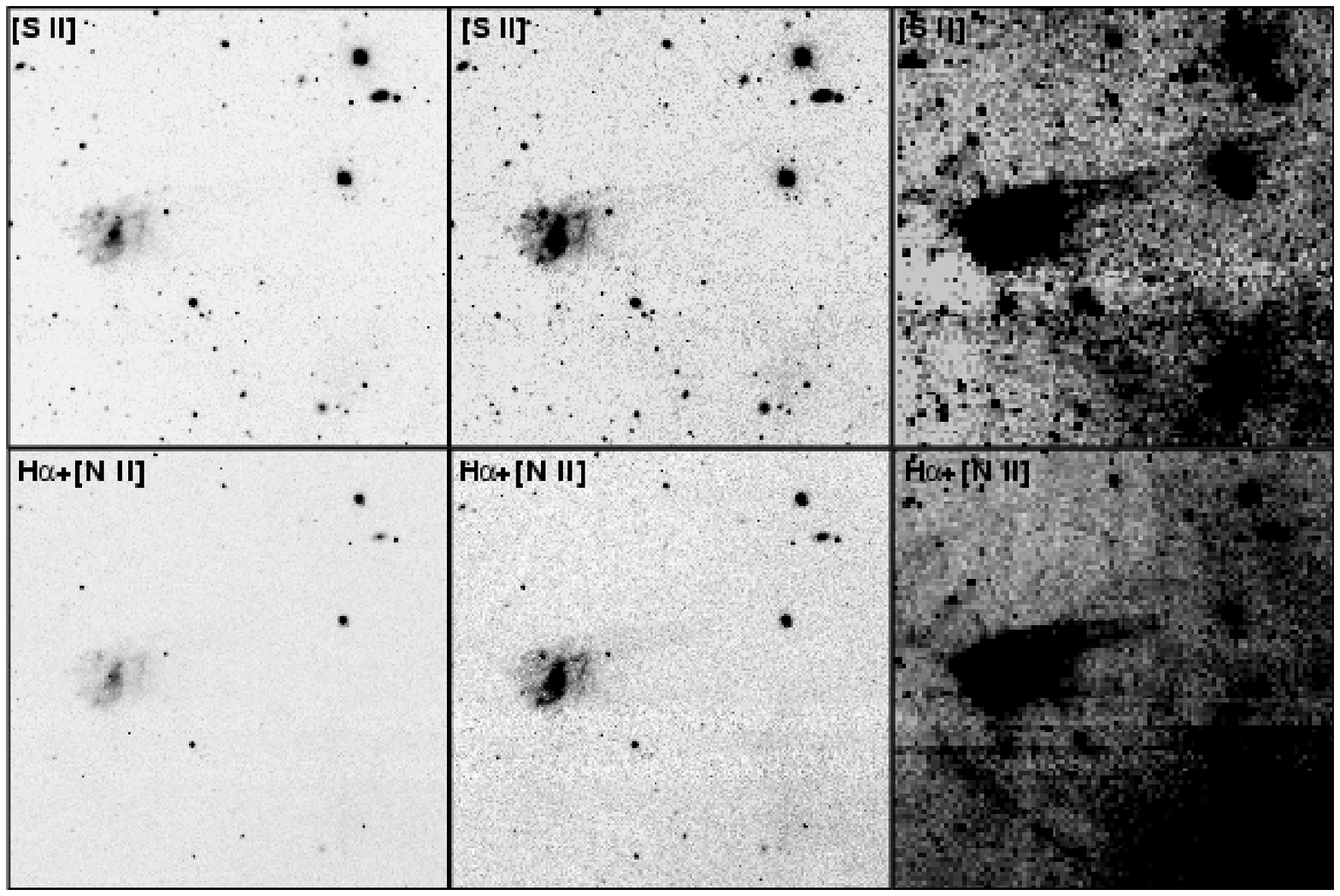}
\caption{
{This series of [S~II] (top) and  \Ha +[N~II] (bottom) images shows a 270\arcsec x270\arcsec\ region around the 
feature NE object (left middle of each panel) denoted in Figure 17. The same image is depicted at different levels of saturation in order to
show the structure at all levels. The right-hand images were made by averaging the original images over 10x10 pixels,
as described in the text. We see that the inner structure is surrounded by a bow-shock with a chord of about 47\arcsec\ and this
is surrounded by a faint thin bow-shock with a chord of 142\arcsec.}
} \label{fig-18}
\end{figure}

In Figure 17 we see that there is extended emission on both the east and west side.
The easternmost extension is an open parabolic form with PA=81\arcdeg, which appears equally well in both our \Ha\ and
[S~II] images and was recorded earlier in \Hb\ (O1998) and \Ha\ (Kwitter, Chu, \&\ Downes 1993). The furthest
east component extends beyond our cutoff at 892\arcsec\ from the central star. 
O1998 records this as extending in \Hb\ to 1190\arcsec\ from the central star.
The west side is a much broader
feature, containing numerous cusps and having a nearly opposite symmetry axis of PA$\approxeq$256\arcdeg. It goes beyond our 866\arcsec\ 
limit, O1998 indicating an extension to 880\arcsec\ and his Figure 3 (bottom left) shows that the object is a series
of well-defined shocks.
Like the east-outer feature , this west-outer feature is very similar in both \Ha\ and [S~II], although the two sides are
quite different in their internal structure. The east-outer feature looks like a series of co-aligned
large shocks, whereas the west-outer feature looks like a series of small shocks having a global
alignment in their average orientation. Borkowski (1993) has argued that the east-outer feature is due to compression 
arising from the Helix Nebula moving into the ambient interstellar medium. This interpretation is tempting
as the measured proper motion for the central star is about in this
direction. Cudworth (1974) reports a photographic determination of 40 mas yr$^{-1}$ towards 73\arcdeg\ and
more recent CCD observations give 33 mas yr$^{-1}$ towards 87\arcdeg\ (H. Harris 2001, private communication). Adopting a value  of 36 mas yr$^{-1
}$
means that the nebula is moving with a tangential motion of 36 \kms\ towards about PA=80\arcdeg. A quantitative
assessment of this interpretation awaits a determination of a radial velocity of the object.

The brightest outer feature is the northeast-arc, which was first reported by Malin and imaged many times
since (Walsh \&\ Meaburn 1987), Kwitter, Chu, \&\ Downes 1993, O1998). It has a characteristic distance from
the central star of 676\arcsec\ (O1998).
Walsh \&\ Meaburn (1987) have compared the velocity of a sample of the northeast-arc nearly east of the 
central star and found that it has two components with an average velocity blueshifted 50 \kms\ with respect
to the main body of the nebula.  
Since the energy of a moving hydrogen atom is
0.005184 x V$\rm ^{2}_{ km~s{^{-1}}}$ eV, then the total energy must be at least 13.0  eV, quite adequate to produce a
collisional ionization of hydrogen 
if there is also a tangential velocity component while the observed velocity is well in excess of the sound velocity, so that it can
easily produce a shock.
The northeast orientation of this shock argues that
it is the result of a large-scale high velocity gas moving into a slower or even static cloud of older material
that has been generally compressed by interaction with the ISM on the side facing the proper motion of the nebula.
The shock interpretation of the northeast-arc is supported by a comparison of this region
in \Ha\ from the S2002 study and their presentation of the 100 \micron\ IRAS results at comparable resolution
(their Figure 11), where one sees that the northeast-arc is barely visible in the 100 \micron\ dust emission.
Although the higher density of the northeast-arc would produce some of this improved contrast, the
higher gas temperature in shock would also play a significant role. The northeast-arc is much more visible in
the [O~III] filter image in O1998 than in our Figure 5. The difference is probably in the filters used, with
the earlier study's filter passing more continuum arising from atomic processes.  

It is possible that the series of shocks that we call the East Outer Feature in Figure 17 is actually the result of
a collimated outflow from the inner nebula and that the counter-part of the West-Outer Feature is the material that
is northeast of the nebula, including the northeast-arc. Within that interpretation the northeast-arc is an indication of
the material flow to the northeast being slowed through interaction with the ISM. The well defined structure
of the northeast-arc and its closer distance to the star is a strong argument for this.

To the southeast (PA=120\arcdeg) we see a parabolic form (in \Ha +[N~II] only, not [S~II] or [O~III]).
Almost opposite there is a much broader component, again only clear in \Ha +[N~II] which has PA$\approxeq$303\arcdeg.
These forms are aligned almost along the axes of the southeast and northwest plumes discussed in the next
section.

We can compare the outer nebula structure displayed in Figure 17 with Figure 1 of S2002
since they look at the same regions, albeit at much lower spatial resolution. Their Figure 1 shows the \Ha\ only
surface brightness at a resolution of 48\arcsec. We see that our east-outer feature extends to about
1160\arcsec , where it broadens (their B1 feature). They identify a second source (called B2) at about 780\arcsec\ 
from the central star at about PA=265\arcdeg, which our Figure 17 shows is  four objects with the appearance of shocks.
Their Figure 1 also shows that our southeast-outer feature extends to about 1200\arcsec\ (their C1) and that our northwest-outer
feature extends to about 1000\arcsec\ (their C2).  Their feature A1 is our feature ``NE Object'', which is discussed
in \S\ 4.3.3 and their feature A2 is a region of several condensations within the west-outer structure and
appears to have no particular importance when viewed at our spatial resolution. 

S2002 argue that there is an extended \Ha\ glow surrounding the Helix Nebula.  Our observations indicate that
this is composed of collimated flows, rather than being a general cloud. Since shocks are likely to play an important
role in their excitation and we don't know the multiple conditions necessary for calculating the emissivity of a 
shock, we cannot comment on the details of this extended glow.  The shocks may be forming in partially ionized
material that is no longer directly ionized by the central star. The surface brightness in \Hb\ of the northeast and southwest-outer
regions at 520\arcsec\ from the central star is about 3\%\ the surface brightness of the main-ring. 
If this emission comes from photoionized gas in 
a sphere of 50\%\ larger size,  and most of the main-ring emission comes from a disk about 100\arcsec\ in line of sight thickness 
(O1998), then scaling from the density (N$\rm _{e}$) of the main-ring of 60 electrons \cmq\ indicates an outer cloud gas of about
3 electrons \cmq. The recombination timescale for a 10,000 K gas is about 120,000/N$\rm _{e}$ years (Osterbrock 1989). This
means that the unshocked recombination timescale for the outer material is about 40,000 years, which is much longer than
the lifetime of the nebula. This means that this outer material could have been ionized at an earlier stage in the
evolution of the nebula and we see it now as it is fading out as it recombines.

In summary, we can say that there is evidence for several flows of nebular material that has produced the outer structure of
the Helix Nebula. There is a bipolar flow that produces the well defined southeast-outer+northwest-outer features. In addition, there is a series of shocks 
extending towards PA=80\arcdeg\ which probably represents where high velocity collimated flow from the center of the nebula shocks 
against the ambient ISM. Finally, there is a broader bipolar flow along a east-northeast to west-southwest axis, which
is shocked and compressed by interaction on the side confronting the ambient ISM (thus producing the northeast-arc)
and extending farther in the direction away from the systemic proper motion.

\subsubsection{The plumes}

The features we call the northwest and southeast plumes are seen in most well exposed images of the main ring. They are apparent in 
all the emission lines in this study, which means that these are regions being penetrated by radiation from the central star, rather
than being shadowed by the ionization structure of the inner-disk. On the northwest side the plume 
appears to at least partially interrupt the outer-ring whereas on the southeast side the outer-ring is simply fainter where it passes.
The orientation of the plumes is PA=303\arcdeg\ and shares the orientation with the much larger southeast-outer and northwest-outer features
to which there must be a connection. As indicated in the previous section, these features must be an indication of a bipolar flow.
The most likely interpretation (O1998) is that these features are all extended structure perpendicular to the inner-disk
of the nebula.

\subsubsection{The NE Object and the NE shock}

The NE Object, whose position is indicated in Figure 17 was originally discovered by Malin (1982).
He called it object A, reported it as visible only in his red bandpass image, and described it as ``a curious cometlike
object.'' It appears in the O1998 \Hb\ image, but is not discussed. It appears in the S2002 \Ha\ image, where it is 
called object A1. It falls only into our CTIO fields, where it is visible in both the \Ha +[N~II] and [S~II] images, but
not the [O~III] image.  It does not appear in the R band image of Walsh \&\ Meaburn (1987). The region around the NE object
is depicted in Figure 18 in a series of maximum surface brightness images. The inner-most region of the object appears
much like an open spiral galaxy. However, the deeper images begin to show a bow-shock shape around the inner structure,
which rules out the interpretation of its being a galaxy. The orientation of this inner bow-shock is slightly less
than PA=90\arcdeg , indicating that this is a shock between a gaseous feature associated with the Helix Nebula and the
local ISM. The deepest images show an outer bow-shock that is much brighter on its south side, is visible
only in the \Ha +[N~II] image, and has the same orientation as the inner bow-shock. The right-hand images of Figure 18
were made from images averaged over 10x10 pixels in order to increase the signal to noise ratio using the IRAF task 
{\it blkavg}.

\begin{figure}
\plotone{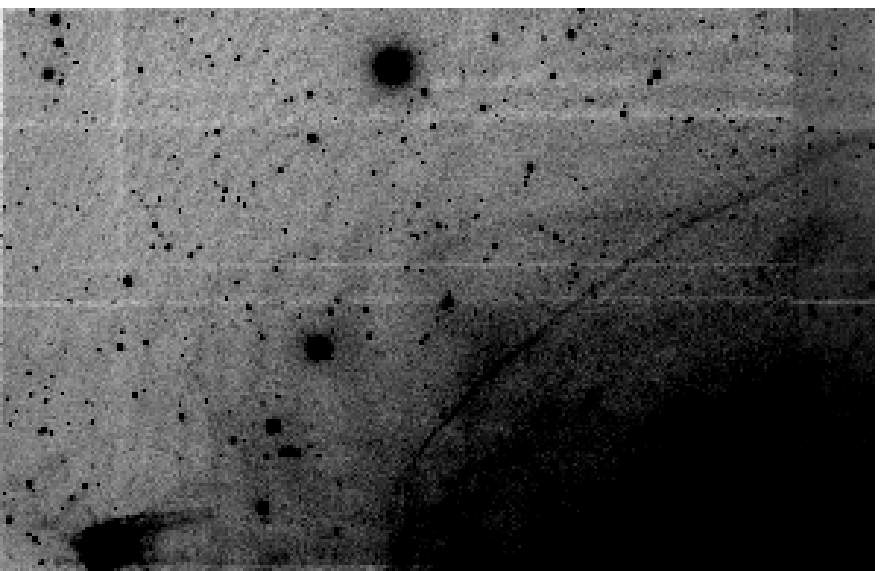}
\caption{
{This \Ha +[N~II] image (789\arcsec x 507\arcsec) of the northeast portion of the outer Helix Nebula shows the 
NE object of Figure 17 in the lower left corner and the much larger, thin bow-shock called the NE shock. The alignment
of the inner and outer bow-shocks associated with the NE object and the NE shock suggests a common origin. The alignment
is indistinguishably the same as the direction of the proper motion of the central star.}
} \label{fig-19}
\end{figure}

\begin{figure}
\plotone{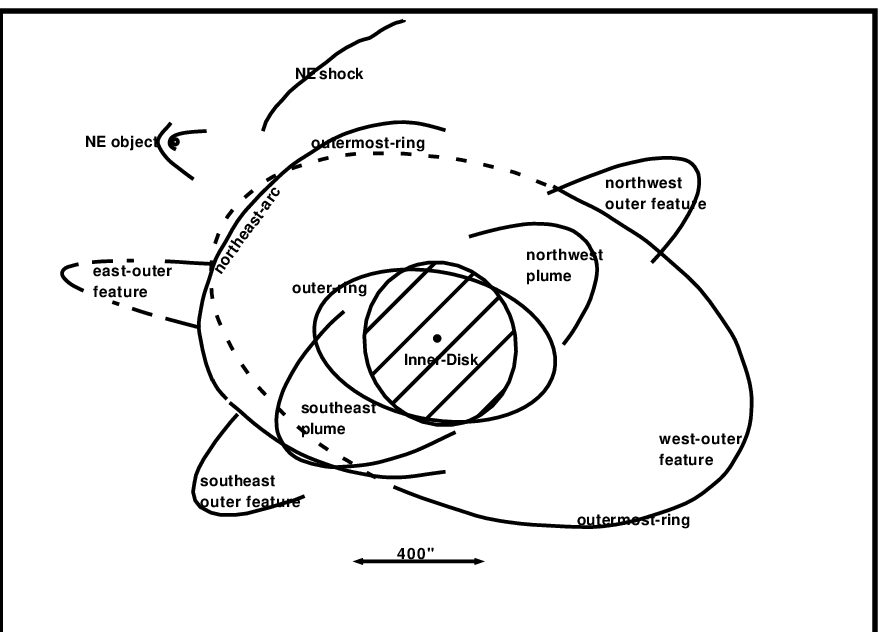}
\caption{
{This drawing depicts the major features of the Helix Nebula as discussed in the text. The outermost-ring is composed of
a single ring whose southwest extension is seen as the west-outer feature and whose northeast extension is compressed
into the northeast-arc.}
} \label{fig-20}
\end{figure}

Immediately to the west of the NE object is the apex of a much larger bow-shock. Figure 19 shows both objects, again
using the 2.64\arcsec\ pixels resulting from averaging the image. This larger bow-shock, which we'll call the NE shock,
is quite thin, being about FWHM=2.6\arcsec\ on the images before averaging. It is much brighter in \Ha +[N~II], only
the brightest part being detectable in [S~II].  It shares the same symmetry axis as the NE object, which means that it
too appears to be caused by interaction of an outflow from the nebula interacting with the ambient ISM. 

The common alignment of both small shocks close to the NE object and the larger NE shock is difficult to explain. It
links the three shocks and argues that they represent some type of intermittent collimated flow as one sees in several
Herbig Haro objects. However, there is no obvious source for such a flow since the symmetry axis passes north of the
north boundary of the main-ring. One can argue that it is not associated with the Helix Nebula, but the agreement of 
alignment with the proper motion makes this unlikely.  

\subsection{The Bars}
The bar features (NE-bar and SW-bar in Figure 6) also merit discussion. These are the brightest portions of 
the main-ring of the nebula, as shown in Figure 1. [O~III] emission is strong in the NE-bar but does not stand out from the main-ring in
the SW-bar region. In \Hb\ the SW-bar is slightly brighter than the NE-bar while in [N~II] the SW-bar is much brighter
than the NE-bar. In [S~II] emission the SW-bar stands out, but is not noticeable above the main-ring in the NE-bar
region. As we see in Figure 3, these bar features occur where the outer parts of the inner-disk and outer-ring
overlap in the plane of the sky, which to the first order can explain why the main-ring is brightest there.  However,
this does not explain the very different ionization differences, with the NE-bar being higher ionization.  

If the inner-disk and the outer-ring were thin and had the same orientation angle in PA, then one would expect that the portions 
of the outer-ring that are shielded from direct ionization radiation from the central star (through the disk being
optically thick) would be low ionization. However, the orientations are not the same. The ellipse fit to the form
of the inner-ring portion of the inner-disk (Figure 3) indicates that the symmetry axis is pointed towards 
288\arcdeg. If the plume features are the result of extended structure perpendicular to the inner-disk, then the
orientation is 303\arcdeg, the 15\arcdeg\ difference being within the range of uncertainty, which is larger for the 
former determination. The orientation of the outer-ring is better determined and is about PA=168\arcdeg. We can adopt
values of 303\arcdeg\ and 168\arcdeg\ for this discussion, this being the PA of the more distant side of the 
symmetry axis.  This difference of orientation means that one can expect varying high and low ionization peaks
depending on the elevation angles involved. The higher ionization NE-bar feature would be the result of high elevation
angle illumination of a higher portion of the far side of the outer-ring, while the lower ionization SW-bar feature
would be the result of looking at a lower elevation region of the near side of the outer-ring. This interpretation
is consistent with the fact that the two bar features are not exactly 180\arcdeg\ apart.

\subsection{A Composite Model for the Helix Nebula}

In Figure 20 we have drawn a composite of the primary features seen in the Helix Nebula. A pattern of associations 
is obvious.  The northwest and southeast plumes are probably extended regions perpendicular to the inner-disk, as 
argued previously (O1998).  The northwest and southeast-outer features are extensions of the plumes. The axis of 
rotation of the inner-disk passes through these features, with the further side having a position angle between
288\arcdeg\ (as argued by the orientation of the inner-ring and the PA-radial velocity pattern) and 303\arcdeg\ (as
argued by the orientation of the plume and associated outer features). The core of this disk is seen in HeII and 
is surrounded by a density enhancement well beyond which 
is a progression of ionization states, with the disk being optically thick to hydrogen LyC in its outermost regions.

The outer-ring lies almost perpendicular to the plane of the inner disk. It has no inner material, is of lower 
ionization and it too becomes optically thick to hydrogen LyC in its outermost region. 
Its [O~III] emission is
what gave rise to the M1998's argument that there was a second [O~III] emitting region within a single torus.

It is notable that the orientation of the outer-ring component of the main-ring of the nebula has a similar orientation
and form as the combined 
northeast-arc and the west-outer features. This suggests that they share a similar structure, with the west-outer
feature and the northeast-arc both being part of another (outermost) ring having an intrinsic diameter of 1700\arcsec , but 
is compressed on the side impinging on the ambient ISM.  
This outermost-ring is almost certainly better described as a torus as the Walsh \&\ Meaburn 1987 spectra indicate
that the spectrum of the eastern portion of the northeast-arc is split into two components.
It is likely that it is a flattened toroidal structure since the surface brightness outside of the shocks is relatively
constant and extends in as far as the outer-ring.
This outermost-ring may in part be collisionally
excited due to the shocks being formed, although some photoionization could be occurring as the result of scattered
LyC photons arising in the plumes and their associated outer features.  The density of the ring could be much more than
that calculated in \S\ 4.3.1 since the line of sight thickness would be much less. This outermost-ring is
probably a thin disk, with the greatest densities in the regions being compressed by shocks. It is impossible to 
quantitatively assess the role of shocks without a complete velocity mapping, but the reasonableness of the argument
is supported by the fact that a sphere of 1.76 pc (1700\arcsec) diameter would enclose a total mass of 0.07 M$_{\hbox{$\odot$}}$, if the
ambient density of the ISM is one hydrogen atoms per cubic centimeter. This is probably comparable or greater than
the mass in the outermost ring, so that significant interaction with the ISM would be expected. This radius of the outermost-ring
(0.89 pc) falls well within the size distribution of extended haloes around other PN (Corradi \etal\ 2003). 
The Corradi \etal\ paper also establishes that large extended haloes are not uncommon, being seen in several tens of
percent of all PN. Such extended outer structures can be the result of the first phases of mass-loss from the
precursor central star (Corradi \etal\ 2000).

The nature of the other features remains uncertain.  The east-outer feature appears to be a series of shocks that
originate from the direction of the central star.  The NE object and its associated shocks resembles a mach-disk
with accompanying shocks. Its symmetry axis is similar to that of the east-outer feature, a property shared with the
NE shock feature. However, their position and symmetry axis argues against an origin in the central star, raising the
possibility that they have a common driving source that is not within the main-ring.

\subsection{A multiplicity of axes of the outflows}

Our new composite picture for the  Helix nebula suggests that it is a poly-polar, rather than
bipolar planetary nebula.  There are two axes of symmetry, the inner disk's with
an  axis tilted 23\arcdeg\ from our line of sight and the outer-ring's with an axis tilted
53\arcdeg\ from our line of sight. Moreover, the inner-disk is at the center of the
northwest and southeast plumes which form point symmetric structures with respect to the central star.
Multiple axes and point symmetric structures have been seen more obviously in other
planetary nebulae because we observe them closer to  edge-on.  These poly-polar planetary
nebulae (L\'opez et~al. 2004) include, e.g.: NGC 6302  (Meaburn \&\ Walsh 1980a, 1980b),
NGC 2440 (L\'opez et~al. 1998), KjPn 8 (L\'opez et~al. 2000), J320 (Harman et~al. 2004) and
the class of quadrupolar nebulae described by Manchado, Stanghellini \&\ Guerrero (1996).
The nature of the Helix's multiple axes has previously eluded detection because we see them almost
pole-on and its two axes of symmetry give it the confusing eponymous helical appearance. Of
the other poly-polar planetary nebulae, the Helix model bears the closest resemblance to that of
NGC 2440, except that NGC 2440 is much younger ($\sim$1600 yrs) and is seen almost edge-on.
Both nebulae have strong molecular hydrogen emission suggesting a large reservoir of
neutral gas (although the Helix's observed \htwo\ is mostly in the knots), an inner-disk with point symmetric features at its poles, an outer-structure
and outer-most structure, all with slightly different axes of symmetry . This comparison
suggests that the outer-ring and outer-most structure may be limb brightened edges of more
cylindrical structures rather than simple rings. Kinematical studies of these poly-polar
planetary nebulae reveal different velocities for the different structures as we also observe in
the Helix. 

The axisymmetric structure of planetary nebulae has been attributed to the dynamical
influence of a companion star or even Jovian planets (e.g. Soker 2001, Soker \&\ Rappaport 2000,
Soker \&\ Rappaport 2001).  The highly collimated jets, such as the northwest and southeast plumes of the
Helix, may have been caused by an accretion disk, probably formed due to the influence of a
binary companion (e.g.  Soker \&\ Rappaport 2001),  and by magnetic fields
(Garc\'{\i}a-Segura et~al. 1999).  The mechanism which creates poly-polarity  is less clear
but many suggest variations on the binary companion theme.   For example, Manchado, Stanghellini \&\ Guerrero (1996)
suggest the influence of a triple star system may also cause poly-polarity.
The point symmetry often observed in poly-polars may be due to the orbital motion
of a binary system (Soker \&\ Rappaport 2001). 

The idea of the nebula being shaped at least in part through being a binary star
is strengthened by the fact that the most plausible interpretation of the central point-like
x-ray source is its being a dwarf M star, as discussed in \S\ 4.1.
Temporal variability in  the
x-rays and in \Ha\ supports this idea (Gruendl et~al. 2001, Guerrero, Chu, \&\ Gruendl 2004). As pointed out by Miranda et~al. (2001) the kinematics of all the
different structures in a poly-polar planetary nebulae may provide detailed
information about the binary star orbital parameters.
The next step in understanding the structure of the Helix and its origin
from a putative binary system will require detailed kinematical data of all the interesting
features that we outline in this paper.

In this study we have shown that the combination of HST-ACS and CTIO images, together with published radial velocity
information, has allowed us to construct a refined model of the Helix Nebula. There have been at least three epochs
of mass loss, corresponding to the inner-disk, the outer-ring, and the outermost-ring, with the latter two arguably having
similar orientations, while the inner-disk is almost orthogonal to them. The outermost-ring shows the effects of
interaction with the ambient ISM and there are numerous features that are apparent shocks. The nature of a few associated outer
features (the NE object and the NE shock) remain unexplained.

\acknowledgments

We gratefully acknowledge the work of many STScI colleagues who contributed to this project. Outstanding among
them are Max Mutchler who took the lead in making the mosaic of the ACS images, Zoltan Levay who accomplished
the difficult task of combining the CTIO and ACS images, and Lisa Frattare who helped make and reduce the
CTIO images.

This work was supported in part by an STScI grant for the archive program GO 9944 and by the internal STScI funds, DDRF D0001.82319.

Facilities: {\it HST(ACS)}, {\it CTIO(MOSAIC)}.

\clearpage
\clearpage


\clearpage

\end{document}